\documentclass[12pt,english]{article}
\usepackage[T1]{fontenc}
\usepackage[latin9]{inputenc}
\usepackage{geometry}
\usepackage{amsmath}

\usepackage{amsfonts}
\usepackage{amssymb}
\usepackage{graphicx}
\usepackage{color}
\usepackage{float}

\makeatletter
\newcommand{\lyxaddress}[1]{
	\par {\raggedright #1
	\vspace{1.4em}
	\noindent\par}
}

\makeatother

\usepackage{babel}
\begin{document}

\title{\textbf{Introduction to Regge Calculus for Gravitation}}

\author{R.R. Cuzinatto, C.A.M. de Melo and C. Naldoni de Souza}

\date{}

\maketitle

\lyxaddress{Instituto de Ci\^encia e Tecnologia, Universidade Federal de Alfenas \\ BR 267 -- Rodovia Jos\'e
Aur\'elio Vilela, 11999, CEP 37701-970, Po\c cos de Caldas, MG, Brazil}

\begin{abstract}
With the theory of general relativity, Einstein abolished the interpretation
of gravitation as a force and associated it to the curvature of spacetime.
Tensorial calculus and differential geometry are the mathematical
resources necessary to study the spacetime manifold in the context
of Einstein's theory. In 1961, Tullio Regge published a work on which
he uses the old idea of triangulation of surfaces aiming the description
of curvature, and, therefore, gravitation, through the use of discrete
calculus. In this paper, we approach Regge Calculus pedagogically,
as well as the main results towards a discretized version of Einstein's
theory of gravitation.
\end{abstract}

\section{Introduction}

Since the ancient Greeks, the method of decomposing a complicated
problem in simpler parts is one of the fundamental pillars of science
development. Indeed, it is possible to consider this abstraction as
one of the techniques of logical thought since it permeates natural
science including mathematics.

Democritus (c.460 BC) introduced the idea of decomposing a complex object into
fundamental indivisible and smaller pieces, and it took almost 2,500 years for this proposal 
being coherently implemented by Quantum Mechanics. We can find a similar type of reasoning
in the efforts of Eudoxus (408-255 BC) in his efforts to calculate
areas using the \emph{Method of Exhaustion}\footnote{The method of exhaustion
is a way of solving the problem of squaring the circle by building a
circumference through infinite small line segments.}.
The exhaustion method has a relation to the \emph{Finite Element Method} and the graphics computational methods used to smooth out surfaces.




Regge Calculus \cite{Re61,Mi17} is an additional element of this set of
discretization methods applied to the description of space-time. Tullio
Regge's ideas were to build the smooth spacetime manifold without using
coordinates. Instead, he used basic concepts of topology. Once these concepts
are more familiar in three dimensions, Regge uses the method of ``Euclidianization'',
where some geometrical quantities of the theory assume complex values
\footnote{In terms of coordinates, this is equivalent to an ``imaginary time''
$x^{4}=ict$.}. Thereby, he demonstrated the results in two or three
dimensions from which the conclusion could be generalized to four
dimensions.This strategy is pedagogically resourceful and will be used in this paper.

\begin{figure}[h]
\begin{center}
\includegraphics[width=7cm]{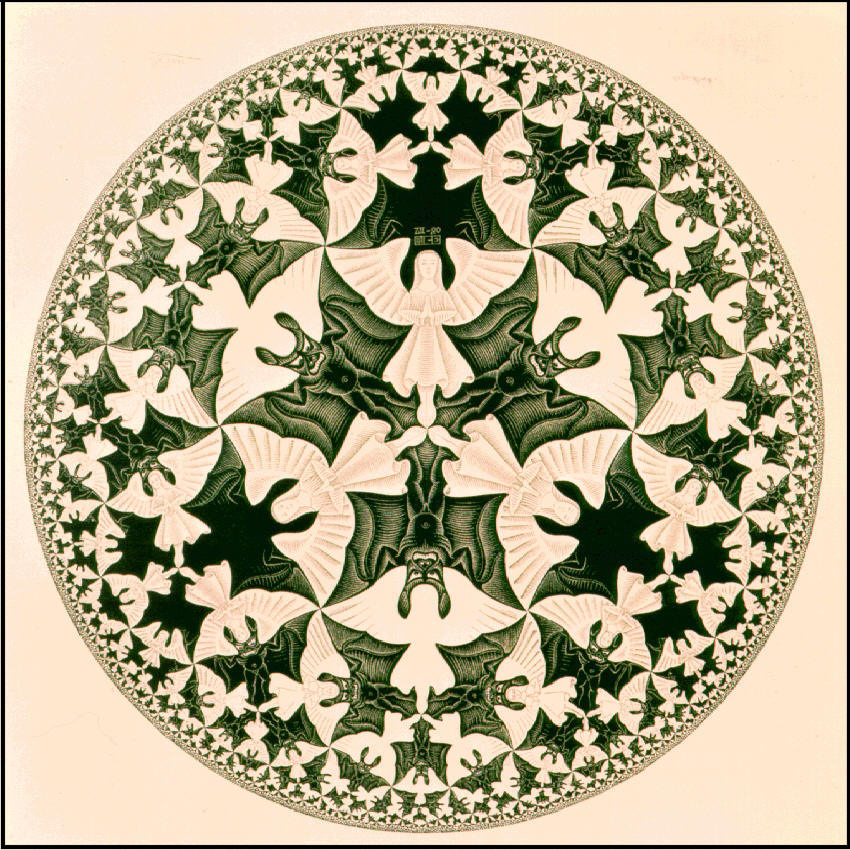}
\end{center}
\caption{The work \textquotedblleft  Circle Limit IV: Heaven and Hell\textquotedblright\ by M.C. Escher (dated 1960) exemplifies the overlay of a surface by mosaics with different characteristics.}%
\label{FigEscher}%
\end{figure}

The same way as the basic elements of matter receive a special name -- atoms --,
there is a denomination for the fundamental elements of geometry in Regge Calculus:
\emph{simplexes}. A simplex is the spacetime manifold fundamental building-block. For
example, think of a two-dimension surface like a wall. The simplexes would be the
tiles or mosaics used to cover it. There is no need for the tiles to be of
the same form or size (contrary to what commonly happens to the usual tiling
in a house), but they should match like the pieces in a puzzle, they should be self-joining in a way that covers the whole surface. Maurice Escher masterfully illustrated this reasoning, e.g. Fig. \ref{FigEscher}.


\section{The Discretization of Space \label{sec-triangulacao}}

Einstein's equations for gravitation supply a systematic way to determine the
geometry of spacetime, which is generally curved. Given a particular
distribution of matter described by the energy-momentum tensor, it is possible in principle
to calculate the independent components of the metric by solving a nonlinear
system of coupled differential equations. In fact, it is only possible to
obtain the solutions analytically when the degree of symmetry of matter distribution is
high. This fact restricts vertiginously the collection of solutions we
have access to without calling upon numerical resources. This difficulty motivates the search for an
alternative method to general relativity to describe the curvature of spacetime.

\begin{figure}[h]
\begin{center}
\includegraphics[width=6cm]{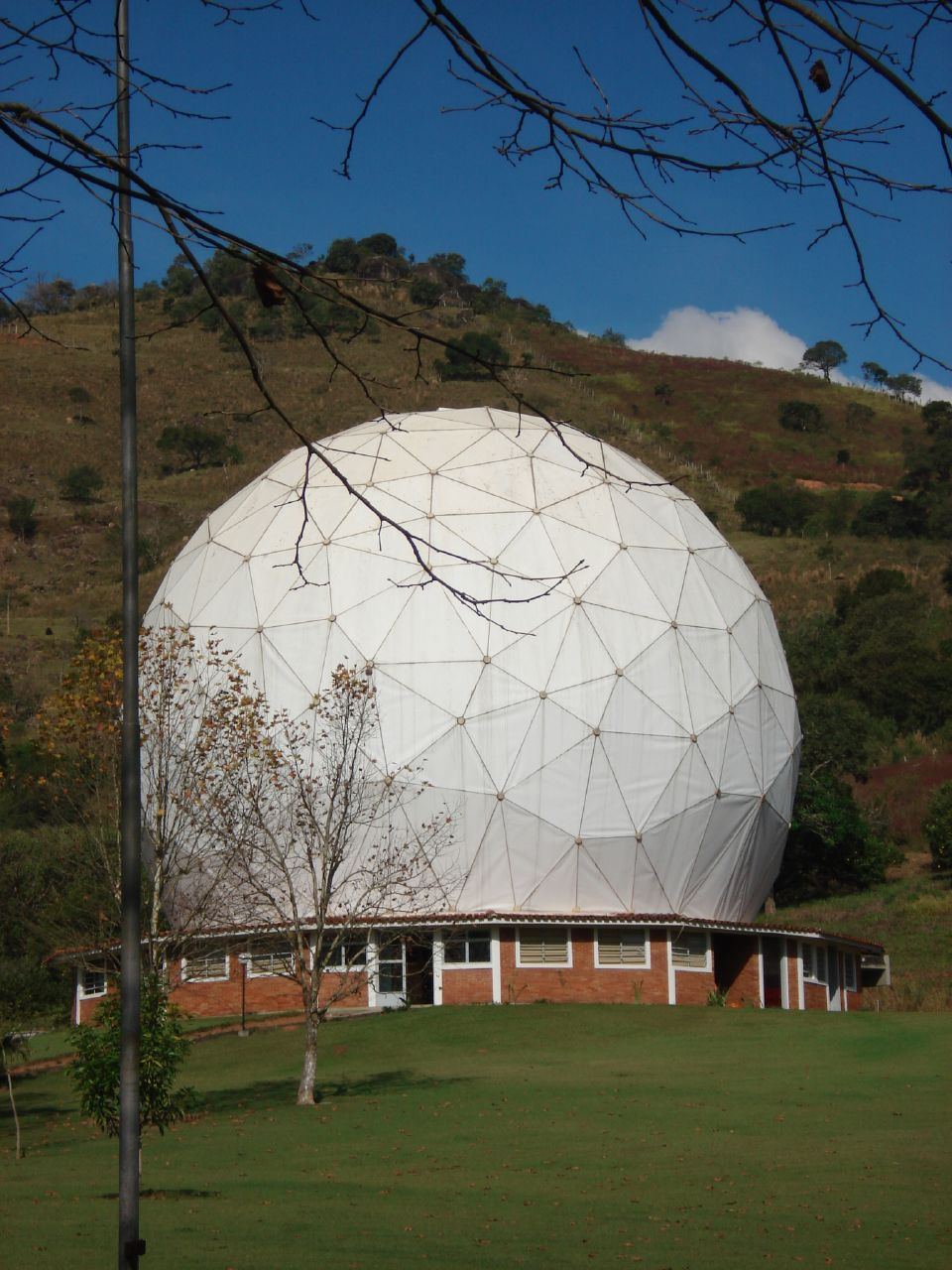}
\end{center}
\caption{The dome protecting Itapetinga $13.4\operatorname{m}$ radio telescope at Atibaia (Brazil) is an example of discretization of a curved surface by a set of juxtaposed plane polygons.}%
\label{FigRadio}%
\end{figure}

Regge's work \cite{Re61} lays down such an alternative, even
though his motivation might have been the solution to mathematical problems in
areas such as in topology, homology, holonomy, and homotopy \cite{Na03}. Regge proposes the discretization of a continuous and smooth manifold into Euclidian simplexes (polyhedrons). The triangulation of the manifold carries a similarity with the homology methods (as in Ref. \cite{Al95}%
,\ Chapter 2). Fig.~\ref{FigRadio} is consistent with this
scenario: The picture shows a hemispherical dome that protects
Atibaia's radio telescope in Brazil; the dome is composed of a multitude of plane triangles connected
edge to edge and vertex to vertex.


The surface triangulation starts by choosing the base simplex (the fundamental shape for the covering polygons). The idea is to take the polyhedrons as similar as possible to regular simplexes (of equal
edges). However, it is not possible to maintain all the sides with the same
length, because we need to accommodate some degree of freedom to fit the curved surface.
The number of required simplexes depends on the magnitude of the surface's curvature. The
more intensely the surface bends, the greater is the number of simplexes necessary to cover it. Also, the higher the density of simplexes (number of simplexes per unity area), the better is the approximation achieved with the discretization process.


\begin{figure}[h]
\begin{center}
\includegraphics[width=11cm]{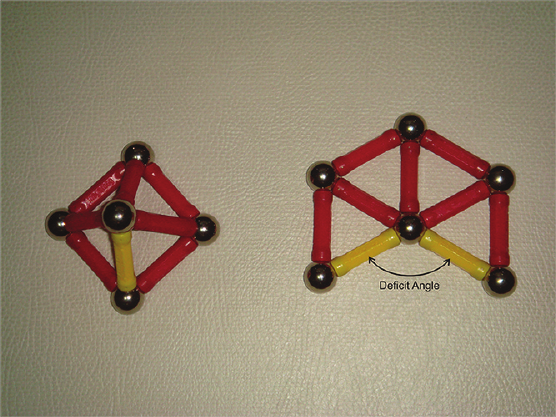}
\end{center}
\caption{The pyramid simplex (3D) -- on the left -- could be cut off along the
yellow edge and be flattened (2D). This way is possible to show the deficit
angle associated with the vertex.}%
\label{FigBrinquedo}%
\end{figure}



Since the elements of the lattice covering the manifold are flat, one might ask: Where is the curvature concentrated? The answer is: Curvature is measured at the vertexes. There is no curvature in between the edges of the adjacent triangles. In fact, consider the point on the tip of a pyramid -- Fig. \ref{FigBrinquedo} (ignore the horizontal basis for the sake of the argument). It is possible to flatten this surface if, and only if, we cut through along one of the edges that go up to the
top $\mathcal{V}$. In this scenario, the sum of the dihedral angles
$\theta_{n}$ (in the $n$ triangles) around $\mathcal{V}$ will not be $2\pi$ in the flatten
surface, as it would be expected in the plane formed of triangles. There will
be an angular difference $\varepsilon$,%
\[
\varepsilon=2\pi-\sum_{n}\theta_{n} \, ,
\]
measuring the curvature of the pyramidal surface. Note that we do not measure any deficit angle when traversing adjoint triangles on the flattened surface except when crossing the edge along with we cut off till the apex, showing that the vertex represents the curvature.

Incidentally, there are many vertexes in a complex surface, each one with its associated angular
deficit $\varepsilon$ characterizing the local curvature. It is in this general surface that we start our quantitative study.


\section{Curvature \label{sec-curvatura}}

There are several polyhedrons around each one of the $M$ vertexes of a discretized manifold.
The number of polyhedrons is large if the magnitude of the local curvature is large. These
polyhedrons touch one another, and the touching edges form a bundle
of many parallel edges -- or \textbf{bones}. This bundle is a \textbf{joint}, designated by the letter $p$; there is a large number  $P$ of joints throughout the manifold. These $P$ joints have a \textit{average} bone density $\varrho$.


\begin{figure}[h]
\begin{center}
\includegraphics[scale=0.5]{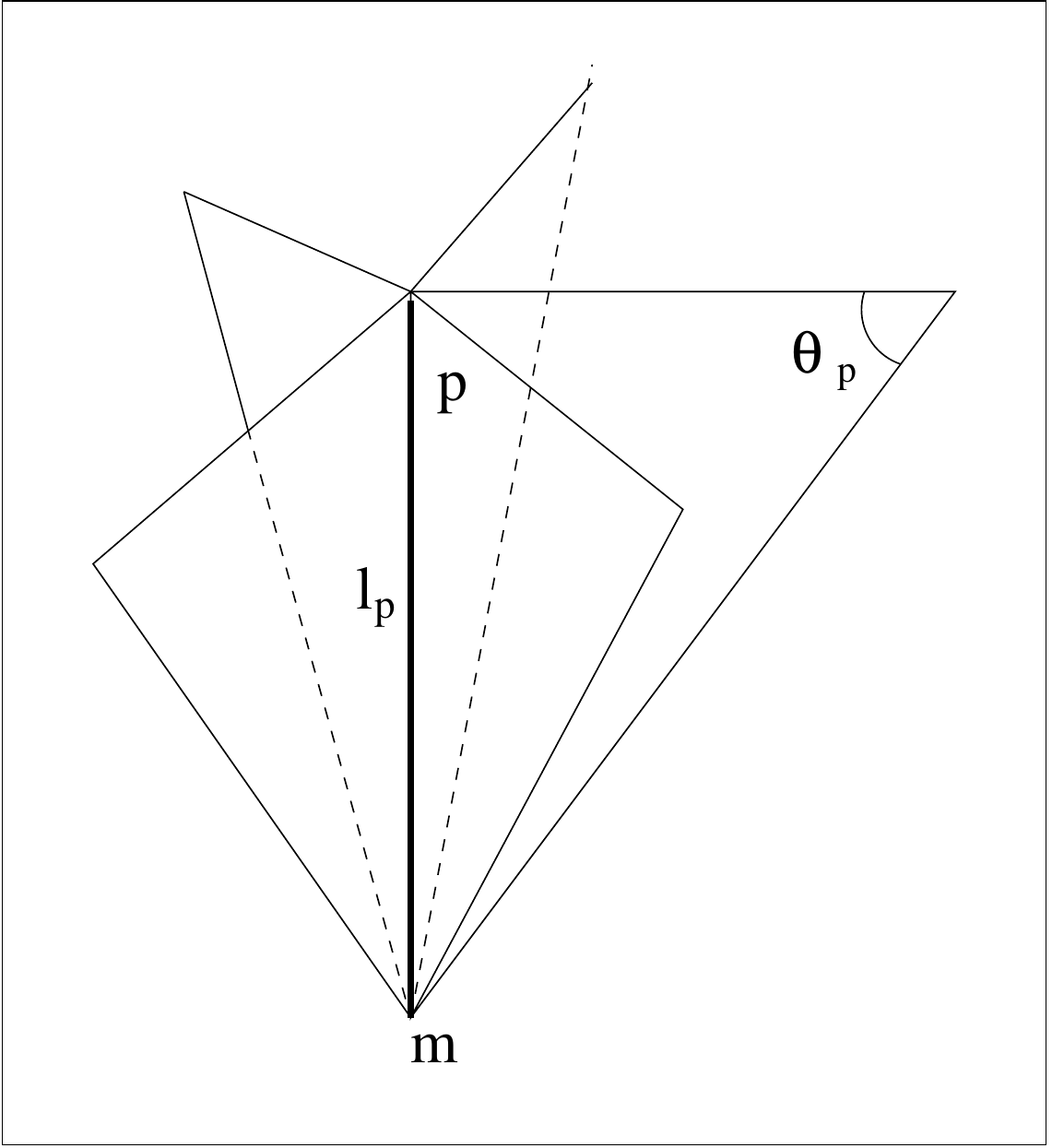}
\end{center}
\caption{A particular joint $p$: bundle composed of $r$ edges ($r=5$). }%
\label{FigPjunta}%
\end{figure}


Fig.~\ref{FigPjunta} shows a specific joint $p$. There is a related bundle of bone of density
$\varrho_{p}$ oriented by the unity vector $\mathbf{U}%
^{p}\left(  \equiv\mathbf{U}_{p}\right)  $ which is parallel to a member of the bundle
and points to the vertex, i.e.,
\begin{equation}
\mathbf{U}^{p}=U^{p}~\mathbf{n}_{p} \, ;~\ \ \ U_{\mu}^{p}U_{p}^{\mu}=1 \, .\label{U}
\end{equation}
We define the components of $\mathbf{U}^{p}$ with respect to a Cartesian
coordinate system fixed in the manifold. This choice of reference frame is always
possible because the hyper-surface is a piecewise-Euclidean manifold.

\begin{figure}[h]
\begin{center}
\includegraphics[scale=0.6]{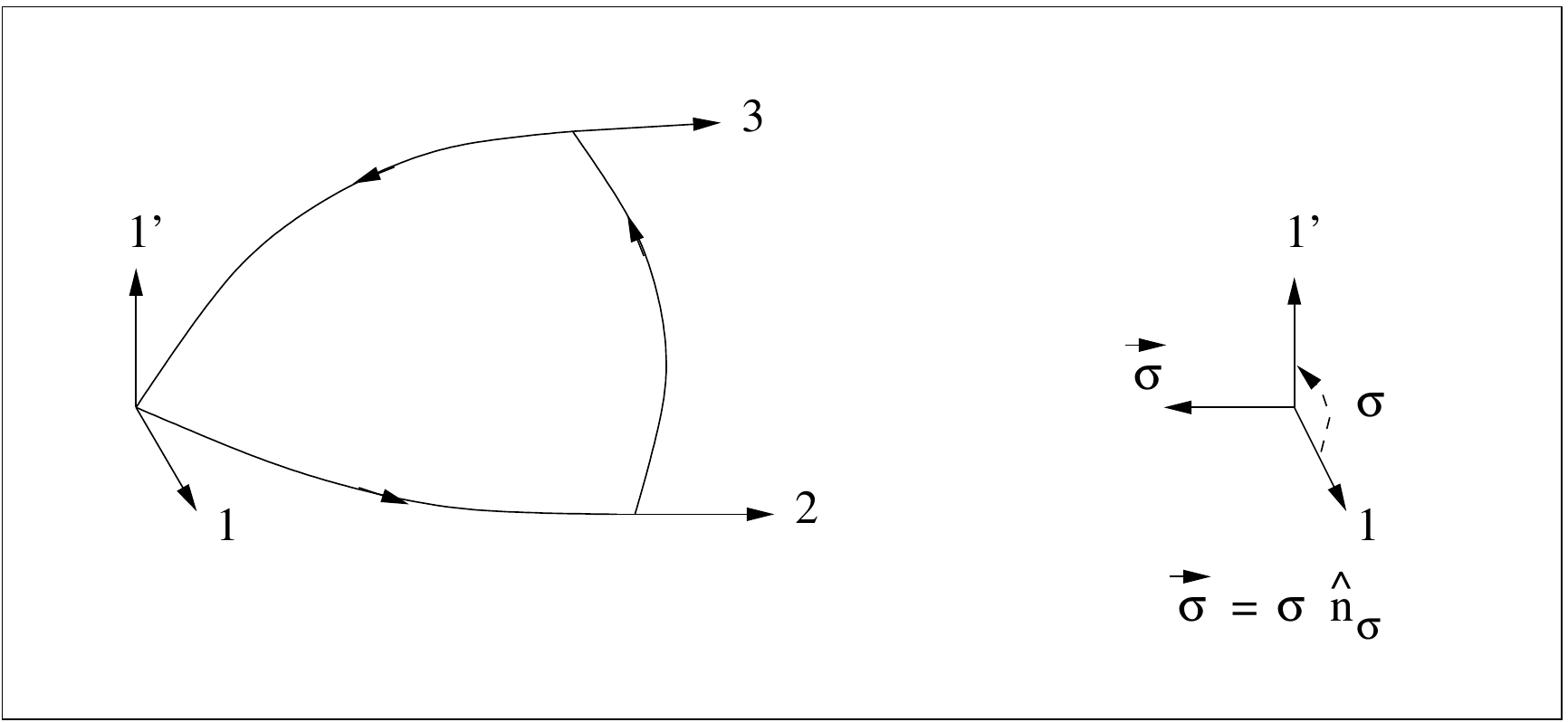}
\end{center}
\caption{Parallel transport of $\mathbf{A}$ in a Riemannian manifold with the approximate shape of a quarter hemisphere (on the left). The vector goes from position $1$ to position $1^{\prime}$ after the displacement $1\rightarrow2\rightarrow3\rightarrow1^\prime$. Note  $\mathbf{A}$ appears rotated with respect to its original position (as indicated on the right): the rotation angle is $\sigma$; $\vec{\sigma}$ is the associated vector built from the unity vector $\hat{n}_p$ which is ortogonal to the plane containing $1$ and $1^\prime$. (Adapted from Ref.~\cite{Sabbata}.)}%
\label{FigTransp}%
\end{figure}

According to the discussion in Sect.~\ref{sec-triangulacao}, the
curvature in joint $p$ is concentrated at the vertex $m$ and the deficit
angle $\varepsilon_{p}$ quantifies it. In the case of continuous manifolds of General Relativity, the
Riemann tensor $R_{~\nu\alpha\beta}^{\mu}$ quantifies spacetime curvature\footnote{In fact, there is a language abuse here: curvature is a quantity associated with the connection $\Gamma_{~\alpha\beta}^{\mu}$ defined over the manifold. In General Relativity the torsion is zero, and the
only feature of $\Gamma$ is the curvature built from it. The curvature is then said to be a characteristic of the spacetime itself.}. It is necessary to study the parallel transport of a vector $\mathbf{A}$ along
an infinitesimal closed loop\footnote{The loop is always finite (but not infinitesimal) in the discretized manifold; as a consequence, the measurement of surface's curvature is
non-local.} in order to relate the discretized manifold $\varepsilon_{p}$ with the continuous manifold $R_{~\nu\alpha\beta}^{\mu}$. A loop in a continuous (Riemannian) manifold is represented in Fig.~\ref{FigTransp}.


Vector $\mathbf{A}$ rotates with respect to its initial position,
throughout the process of being transported. This rotation is the disclination property of the curved space\footnote{In a Weitzenb\"{o}ck
manifold of non-zero torsion, vector $\mathbf{A}$ would appear displaced with
respect to the initial position after the parallel transport. The loop never closes,
and the \textbf{displacement} property is quantified not by an angle but by a
vector, the Burgers' vector $\mathbf{b}$ \cite{Oz12}.}.


\begin{figure}[H]
\begin{center}
\includegraphics[scale=0.6]{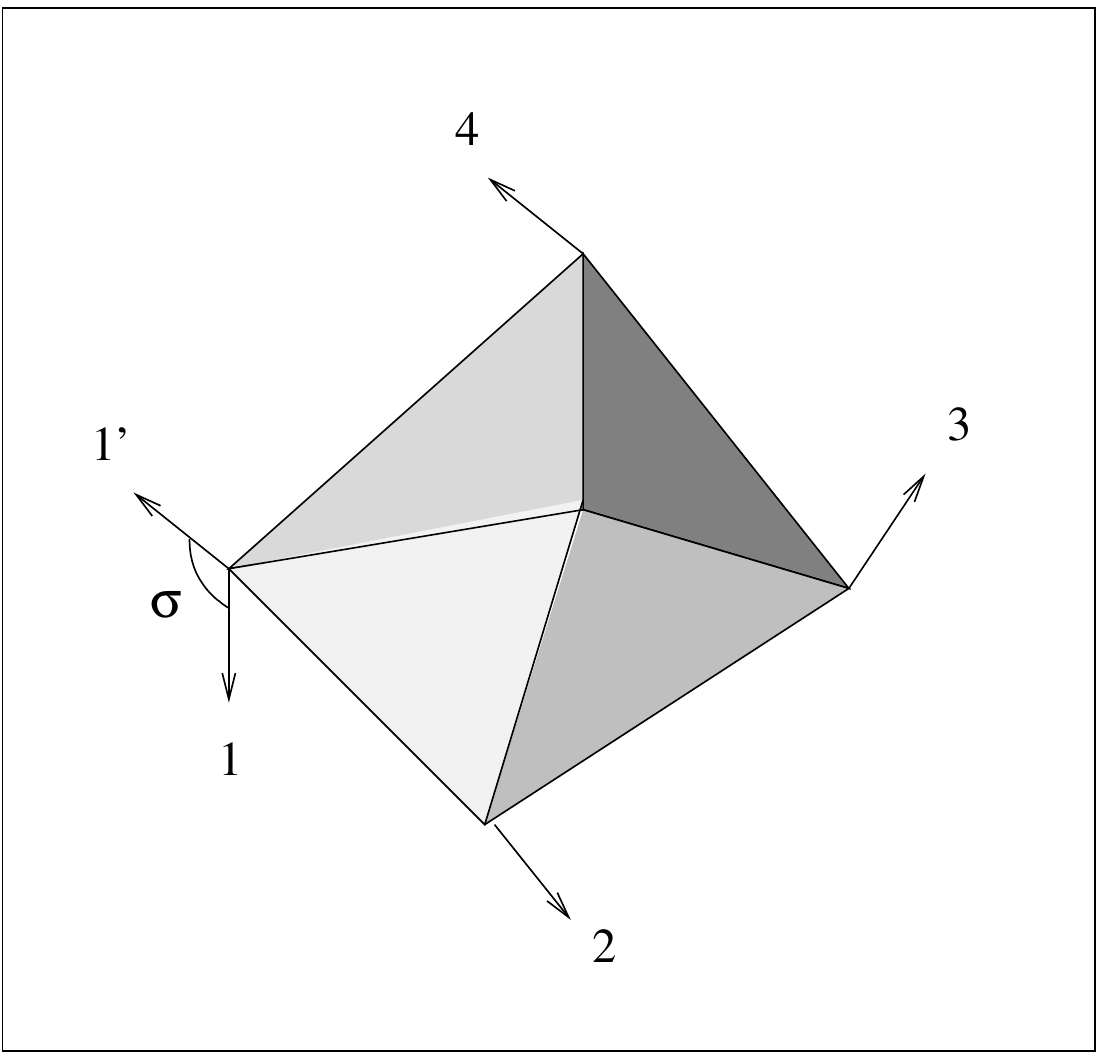}
\end{center}
\caption{Parallel transport of a vector around a set of simplexes.}%
\label{FigDiscreto}%
\end{figure}


The loop in our discretized (simplex) case is taken to be, say, around the bundle
$p$. Let $\Sigma$ be the area enclosed by this loop. If $\mathbf{n}_{\Sigma}$ is
an unity vector ortogonal to this area, then
\begin{equation}
\mathbf{\Sigma}=\Sigma~\mathbf{n}_{\Sigma} \label{area loop}%
\end{equation}
is the area vector associated to that path. The closed path goes around the simplex
polyhedrons that gather at the bundle $p$, cf. Fig.~\ref{FigDiscreto}. At the end of its displacement along the path $1\rightarrow2\rightarrow3\rightarrow4\rightarrow1^\prime$, vector $\mathbf{A}$ rotates by an angle
$\mathbf{\sigma}$,%
\begin{equation}
\vec{\sigma}=\sigma~\mathbf{n}_{p}=\sigma~\mathbf{U}_{p} \, , \label{sigma}%
\end{equation}
pointing along the bundle of bones around which $\mathbf{A}$ rotates. Vetor $\mathbf{A}$ turns to:
\[
\mathbf{\bar{A}}=\mathbf{A}+\delta\mathbf{A} \, .
\]


\begin{figure}[H]
\begin{center}
\includegraphics[scale=0.6]{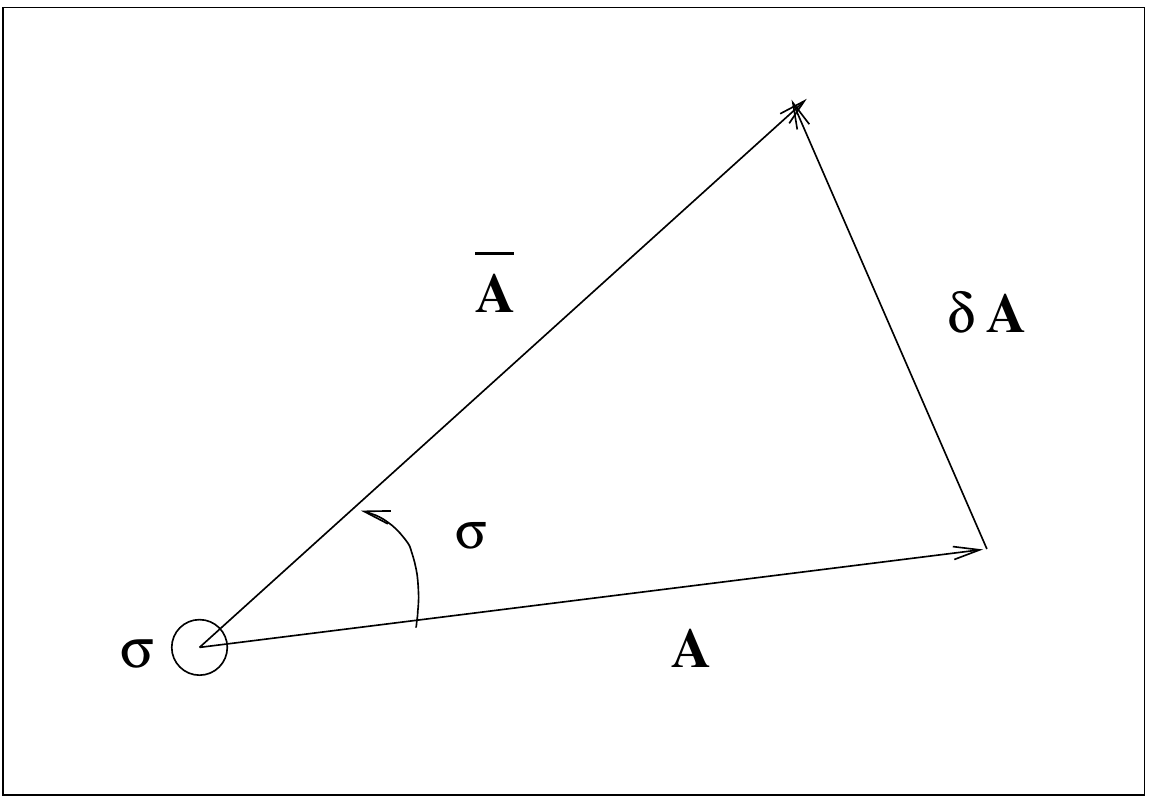}
\end{center}
\caption{Rotation of $\mathbf{A}$ after the parallel transport around a loop in the discretized manifold produces vector $\bar{\mathbf{A}}$. Vector $\mathbf{\sigma}\equiv \vec{\sigma}$ is pointing outward the plane of the page containing both $\mathbf{A}$ and $\bar{\mathbf{A}}$; it gives direction to the rotation angle $\sigma$.}%
\label{FigRotacao}%
\end{figure}


Since the rotation is \textquotedblleft infinitesimal\textquotedblright, the \textquotedblleft arc\textquotedblright\ $\delta\mathbf{A}$ has magnitude $\delta A=\sigma A$. This is displayed in Fig.~\ref{FigRotacao}. The associated vector is\footnote{See e.g. Ref.~\cite{Marion}, Section 1.15.}:
\begin{equation}
\delta\mathbf{A}=\mathbf{\sigma}\times\mathbf{A} \, .\label{delta A sigma}%
\end{equation}
Angle $\sigma$ must be directly proportional to the curvature which is
described by the deficit angle $\varepsilon_{p}$. The proportionality constant
is precisely the number of bones embraced by the loop,  $N$:
\begin{equation}
\sigma=N~\varepsilon_{p} \, . \label{sigma deficit}%
\end{equation}
The reason for that was mentioned before: The higher the number of simplexes $N$ associated to
the vertex $p$ the more significant the curvature and the higher the angular displacement of
$\mathbf{A}$.

\begin{figure}[H]
\begin{center}
\includegraphics[scale=0.6]{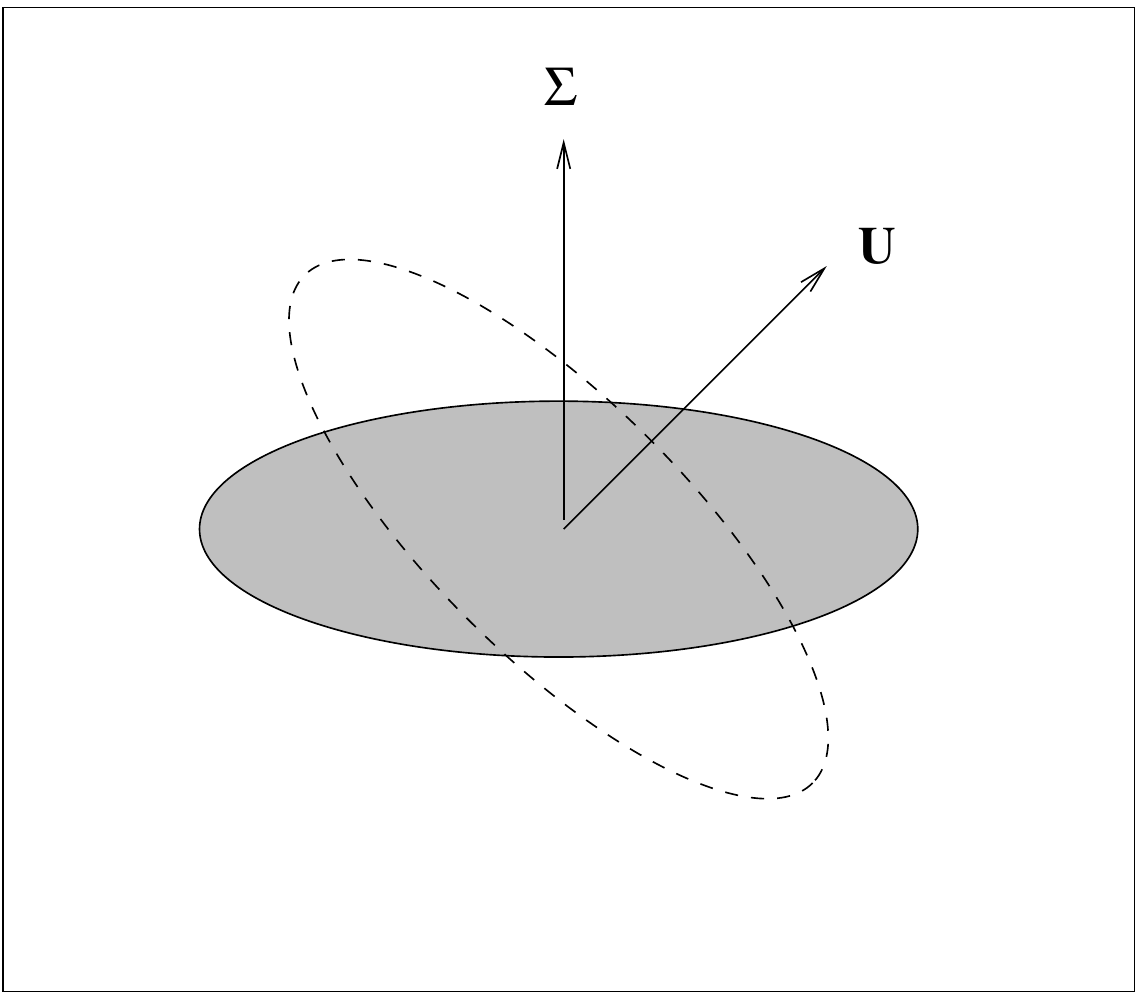}
\end{center}
\caption{Projection of the oriented area $\mathbf{\Sigma}$ onto the direction of vector $\mathbf{U}$ pointing along the bundle of bones.}%
\label{FigProjecao}%
\end{figure}

By the way, $N$ is the product of the density of bones in the joint,  $\varrho_{p}$, by the area resulting from the projection of $\mathbf{\Sigma}$ in the direction of $\mathbf{U}$:%
\begin{equation}
N=\varrho_{p}~\left(  \mathbf{U}_{p},\mathbf{\Sigma}\right)  =\varrho
_{p}~\mathbf{U}_{p}\cdot\mathbf{\Sigma} \, , \label{N}%
\end{equation}
where the symbol $(\,\, ,\,)$ denotes the internal product operation; its is simply the dot product $(\cdot)$ in the context of vector algebra.
(The component of $\mathbf{A}$ along the perpendicular direction to the bundle does not generate contributions to $\vec{\sigma}$.) Fig.~\ref{FigProjecao} sketches the projection achieved by Eq.~(\ref{N}).


By substituting (\ref{sigma}), (\ref{sigma deficit}) and (\ref{N}) into (\ref{delta A sigma}), we obtain the change $\mathbf{A}$ suffered under the parallel transport in the discretized manifold:%
\begin{equation}
\delta\mathbf{A}=\varrho_{p}~\varepsilon_{p}~\left(  \mathbf{U}_{p}%
\cdot\mathbf{\Sigma}\right)  ~\left(  \mathbf{U}_{p}\times\mathbf{A}\right) \, .
\label{delta A}%
\end{equation}
This equation can be expressed in terms of vector components\footnote{The Greek
indexes $\alpha$, $\beta$, $\gamma$, etc. refer to a Cartesian coordinate system defined in a Euclidian hyperplane within the manifold.}:
\begin{equation}
\delta A^{\mu}=\varrho_{p}~\varepsilon_{p}~\left(  U_{p}^{~\nu}\Sigma_{\nu
}\right)  ~\left(  \epsilon^{\mu\alpha\beta}U_{~\alpha}^{p}A_{\beta}\right) \, ,
\label{delta A componentes}%
\end{equation}
where $\epsilon^{\mu\alpha\beta}$ is the totaly antisymmetric Levi-Civita symbol (or permutation
symbol):
\begin{align}
\epsilon^{123}  &  \equiv+1 \, ;\nonumber\\
\epsilon^{\alpha\beta\gamma}  &  =\epsilon^{\gamma\alpha\beta}=\epsilon
^{\beta\gamma\alpha} \, ;\label{Levi-Civita}\\
\epsilon^{\alpha\beta\gamma}  &  =-\epsilon^{\beta\alpha\gamma}=-\epsilon
^{\alpha\gamma\beta} \, .\nonumber
\end{align}
A compact realization of all the features in Eq.~(\ref{Levi-Civita}) is:
\begin{align}
\epsilon_{\alpha\beta\gamma}  &  =\delta_{\alpha}^{1}\delta_{\beta}^{2}%
\delta_{\gamma}^{3}+\delta_{\beta}^{1}\delta_{\gamma}^{2}\delta_{\alpha}%
^{3}+\delta_{\gamma}^{1}\delta_{\alpha}^{2}\delta_{\beta}^{3}+\nonumber\\
&  -\delta_{\alpha}^{1}\delta_{\gamma}^{2}\delta_{\beta}^{3}-\delta_{\gamma
}^{1}\delta_{\beta}^{2}\delta_{\alpha}^{3}-\delta_{\beta}^{1}\delta_{\alpha
}^{2}\delta_{\gamma}^{3} \, . \label{Levi delta}%
\end{align}
Then,%
\begin{equation}
\epsilon_{\alpha\beta\gamma}\epsilon^{\mu\nu\gamma}=\left(  \delta_{~\alpha
}^{\mu}\delta_{~\beta}^{\nu}-\delta_{~\beta}^{\mu}\delta_{~\alpha}^{\nu
}\right)  \, . \label{Levi contracao}%
\end{equation}
Note that we are following Regge's reasoning \cite{Re61} and we treat the
problem in a tridimensional manifold.

$\mathbf{U}_{p}$ components can be conveniently put into the dual form\footnote{The
dual map $(\ast)$ is an operation in which we apply Levi-Civita symbol to the
components of a vector or tensor field $\mathbf{F}$ to build a 
quantity $^{\ast}\mathbf{F}$ with the following feature. The quantity $^{\ast} \mathbf{F}$ has a complementary
number of indexes to the original object $\mathbf{F}$, i.e., the rank of $^{\ast} \mathbf{F}$ is the number of dimensions of the space minus the rank of
$\mathbf{F}$. In this way, the dual of the components of the vector field
$U_{\rho}$ (rank equals $1$) in a tridimensional space ($D=3$) is an object
with $(3-1)=2$ indexes, that is, a rank-2 tensor $U_{\rho\sigma}$. The
dualization technique is crucial for the theory of \emph{differential forms}.
Differential forms are used to cast physical quantities in gravitation and field
theories as coordinate-free invariants. Ref.~\cite{Fel97} is an excellent book containing the dualization technique and differential forms.},
\begin{equation}
U_{\rho\sigma}=\epsilon_{\rho\sigma\lambda}U^{\lambda} \, ;~\ \ \ \ \ \ U_{\rho
\sigma}=-U_{\sigma\rho} \, , \label{U dual}
\end{equation}
Moreover,
\begin{equation}
U^{\lambda}=\frac{1}{2}\epsilon_{\rho\sigma\lambda}U^{\rho\sigma} \, . \label{U componente}
\end{equation}
The factor $(1/2)$ was introduced to avoid double counting of the antisymmetric par of contracted indexes. We use Einstein's convention: There is an implicit sum over repeated indexes.

Analogously, the area has a dual form given by:
\begin{equation}
\Sigma_{\nu}=\frac{1}{2}\epsilon_{\xi\zeta\nu}\Sigma^{\xi\zeta} \, .
\label{area dual}
\end{equation}

With Eqs.~(\ref{U componente}) and (\ref{area dual}), we are able to rewrite Eq.~(\ref{delta A componentes}) for $\delta A^{\mu}$ as:
\begin{equation*}
\delta A^{\mu} = \frac{1}{4}\varrho_{p}\varepsilon_{p}\left(  \epsilon^{\rho\sigma\nu
}U_{~\rho\sigma}^{p}\Sigma_{\nu}\right)  \left[  \left(  \epsilon
_{\kappa\lambda\alpha}\epsilon^{\beta\mu\alpha}\right)  U_{p}^{~\kappa\lambda
}A_{\beta}\right] \, ,
\end{equation*}
where we have used the cyclic property of the indexes in $\epsilon^{\mu\alpha\beta}$. Due to Eq.~(\ref{Levi contracao}):
\[
\delta A^{\mu}=\frac{1}{4}\varrho_{p}\varepsilon_{p}\left(  \epsilon
^{\rho\sigma\nu}U_{~\rho\sigma}^{p}\Sigma_{\nu}\right)  \left(  U_{p}
^{~\beta\mu}-U_{p}^{~\mu\beta}\right)  A_{\beta} \, .
\]
Additionally, $U_{\rho\sigma}=-U_{\sigma\rho}$ and Eq.~(\ref{area dual}) lead to:
\[
\delta A^{\mu}=\frac{1}{4}\varrho_{p}\varepsilon_{p}\left(  \epsilon
^{\rho\sigma\nu}U_{~\rho\sigma}^{p}\frac{1}{2}\epsilon_{~\;\;\nu}^{\xi\zeta
}\Sigma_{\xi\zeta}\right)  \left(  2U_{p}^{~\beta\mu}\right)  A_{\beta} %
\]
or
\[
\delta A^{\mu}=\frac{1}{4}\varrho_{p}\varepsilon_{p}\left(  U_{~\xi\zeta}%
^{p}-U_{~\zeta\xi}^{p}\right)  \Sigma^{\xi\zeta}U_{p}^{~\beta\mu}A_{\beta} \, .
\]
Therefore, the parallel displacement in the discretized manifold is:
\begin{equation}
\delta A^{\mu}=\frac{1}{2}\left(  \varrho_{p}\varepsilon_{p}U_{~\xi\zeta}%
^{p}U_{p}^{~\beta\mu}\right)  \Sigma^{\xi\zeta}A_{\beta} \, .
\label{transp simpletico}%
\end{equation}
Now, let's compare this equation with the expression for $\delta A^{\mu}$ in the
continuous (non-discretized) case. 

General relativity tells us \cite{Sabbata} the effect of parallel transporting a vector $A^{\mu}$ in an infinitesimal loop in the Riemannian manifold is: 
\begin{equation}
\delta A^{\mu}=\frac{1}{2}~R_{\xi\zeta\beta}^{~\ \ \ \ \mu}~\Sigma^{\xi\zeta
}A^{\beta} \, . \label{transp continuo}%
\end{equation}

The analogue of the \textbf{curvature tensor} in the discrete manifold is found by comparing 
Eqs.~(\ref{transp simpletico}) and (\ref{transp continuo}):%
\begin{equation}
R_{\xi\zeta}^{~\ \ \beta\mu}=\varrho_{p}\varepsilon_{p}U_{~\xi\zeta}^{p}%
U_{p}^{~\beta\mu} \, . \label{Riemann simpletico}%
\end{equation}
(We raised and lowered the index $\beta$\ in Eq.~(\ref{transp continuo}), according
to the remarks below.)

The indexes of  $R_{\xi\zeta\beta}^{~\ \ \ \mu}$ in the continuous manifold are
raised and lowered with the help of the metric tensor $\mathbf{g}=g_{\mu\nu
}~dx^{\mu}\otimes dx^{\nu}$. For instance, 
\[
R_{\xi\zeta\beta\nu}=g_{\nu\mu}R_{\xi\zeta\beta}^{~\ \;\;\mu} \, .
\]
However, note that in the simplex Euclidian manifold we have:
\[
g_{\nu\mu}\equiv\delta_{\nu\mu} \, .
\]
Then, we know how to write $R_{\xi\zeta\beta\nu}$ in terms of $\mathbf{U}^{p}%
$: $R_{\alpha\beta\mu\nu}=\varrho_{p}\varepsilon_{p}U_{~\alpha\beta}%
^{p}U_{~\mu\nu}^{p}$.

Let us contract the second and the last indexes of $R_{\alpha\beta}^{~\ \ \mu\nu}$ once this is traditionally defined as the \textbf{Ricci tensor}:
\begin{align*}
R_{\alpha\beta}^{~\ \ \mu\beta}  &  =\varrho_{p}\varepsilon_{p}U_{~\alpha
\beta}^{p}U_{p}^{~\mu\beta}=\varrho_{p}\varepsilon_{p}\left(  \epsilon
_{\alpha\beta\rho}U_{p}^{\rho}\right)  \left(  \epsilon^{\mu\beta\sigma
}U_{\sigma}^{p}\right)  =\\
&  =\varrho_{p}\varepsilon_{p}\left(  \delta_{~\alpha}^{\mu}U_{p}^{\sigma
}U_{\sigma}^{p}-U_{p}^{\mu}U_{\alpha}^{p}\right) \, .
\end{align*}
Since $\mathbf{U}^{p}$ was defined as in Eq. (\ref{U}), the above equation reads: 
\[
R_{\alpha\beta}^{~\ \ \mu\beta}=\varrho_{p}\varepsilon_{p}\left(
\delta_{~\alpha}^{\mu}-U_{\alpha}^{p}U_{p}^{\mu}\right) \, .
\]
Therefore, Ricci tensor is:%
\begin{equation}
R_{\mu\nu}=R_{\mu\beta\nu}^{~\ \ \;\;~\beta}=\varrho_{p}\varepsilon_{p}\left(
\delta_{\mu\nu}-U_{\mu}^{p}U_{\nu}^{p}\right) \, . \label{Ricci}%
\end{equation}

Finally, the \textbf{scalar curvature} (or Ricci scalar) is the index contraction of the Ricci:
\[
R=R_{\alpha\beta}^{~\ \ \alpha\beta}=\varrho_{p}\varepsilon_{p}\left(
\delta_{~\alpha}^{\alpha}-U_{\alpha}^{p}U_{p}^{\alpha}\right)  =\varrho
_{p}\varepsilon_{p}\left(  3-1\right) \, ,
\]
i.e.%
\begin{equation}
R=2~\varrho_{p}\varepsilon_{p} \, .\label{R}%
\end{equation}
The above result makes it clear the equivalence between the Ricci scalar and the deficit angle $\varepsilon_{p}$. This establishes the mapping  between the continuous description of curvature and its discrete counterpart.


\section{Bianchi Identities\label{sec-Bianchi}}

Eq.~(\ref{Riemann simpletico}) is a new way of evaluating curvature in the context of a discretized manifold. In this section we study some properties of the novel Riemann tensor which depends on the simplex structures: number density of bones in a particular joint, the deficit angle and the dual to the vector along the bundle of bones. The results discussed here will be useful for obtaining the discretized version of gravity's field equations in the following Section \ref{sec-EOM}.


\subsection{Properties of $R_{\alpha\beta\mu\nu}$\ and the first Bianchi
identity}

Eq.~(\ref{Riemann simpletico}) satisfies desired properties of the Riemann tensor,
\begin{align}
R_{\alpha\beta\mu\nu}  &  =-R_{\beta\alpha\mu\nu} \, ,\nonumber\\
R_{\alpha\beta\mu\nu}  &  =-R_{\alpha\beta\nu\mu} \, ,\label{propriedades R}\\
R_{\alpha\beta\mu\nu}  &  =R_{\mu\nu\alpha\beta} \, .\nonumber
\end{align}
These are features inherited from $U_{\xi\zeta}=-U_{\zeta\xi}$.

Moreover, the cyclic property of the first three indexes in $R_{\alpha\beta\mu\nu}$
(first Bianchi identity),
\begin{equation}
R_{\alpha\beta\mu\nu}+R_{\beta\mu\alpha\nu}+R_{\mu\alpha\beta\nu}=0 \, ,
\label{Bianchi 1 RG}%
\end{equation}
translates to
\begin{equation}
U_{\alpha\beta}U_{\mu\nu}+U_{\beta\mu}U_{\alpha\nu}+U_{\mu\alpha}U_{\beta\nu
}=0 \, . \label{Bianchi 1}%
\end{equation}


\subsection{The second Bianchi identity}

The second Bianchi identity for a (continuous) curved spacetime,
\begin{equation}
B_{\lambda\alpha\beta\mu\nu}\equiv\nabla_{\lambda}R_{\alpha\beta\mu\nu}%
+\nabla_{\alpha}R_{\beta\lambda\mu\nu}+\nabla_{\beta}R_{\lambda\alpha\mu\nu
}=0 \, , \label{Bianchi RG}
\end{equation}
is verified directly from the Riemann tensor expression in term of the
Christoffel connection:
\begin{equation}
R_{\alpha\beta\mu}^{~\ \ \ \nu}=\partial_{\alpha}\Gamma_{~\mu\beta}^{\nu
}-\partial_{\beta}\Gamma_{~\mu\alpha}^{\nu}+\Gamma_{~\lambda\alpha}^{\nu
}\Gamma_{~\mu\beta}^{\lambda}+\Gamma_{~\lambda\beta}^{\nu}\Gamma_{~\mu\alpha
}^{\lambda} \, . \label{Riemann continuo}%
\end{equation}
Eq.~(\ref{Bianchi RG}) contains the covariant derivative operator, which for a
rank-1 tensor $\mathbf{V}$ with components $V^{\mu}$ reads:
\begin{equation}
\nabla_{\mu}V^{\nu}=\partial_{\mu}V^{\nu}+\Gamma_{~\mu\lambda}^{\nu}%
V^{\lambda}~. \label{derivada covariante}%
\end{equation}

The demonstration of the discrete version of the second Bianchi identity is
laborious: it requires two results of Topology established in the next
sub-sections. Due to the facts that the manifold is flat by pieces and the curvature is
concentrated at the vertexes, it is only natural that the local geometric
properties of the smooth manifold are expressed in term of the skeleton topology.


\subsubsection{Homotopy, holonomy and the deficit angle \label{sec-homotopia}}

It is true that after the parallel transport of $\mathbf{A}$ along an ``infinitesimal'' loop
around the $p$-joint the vector appears rotated: $\mathbf{A}\rightarrow
\mathbf{\bar{A}}$. It is also true the loop closes (since torsion is null \cite{Sabbata}) and the curve is Euclidian by pieces (according to Regge's axioms). Then, we can imagine a picture in which we construct this loop by joining successive curves which connect two arbitrary points in any contiguous simplexes in a given $p$-joint. The mathematical concept related to the deformation that takes a given loop into another is called \emph{homotopy} \cite{Al95}.


\subparagraph{Definition (Homotopy)}

\begin{figure}[h]
\begin{center}
\includegraphics[scale=0.8]{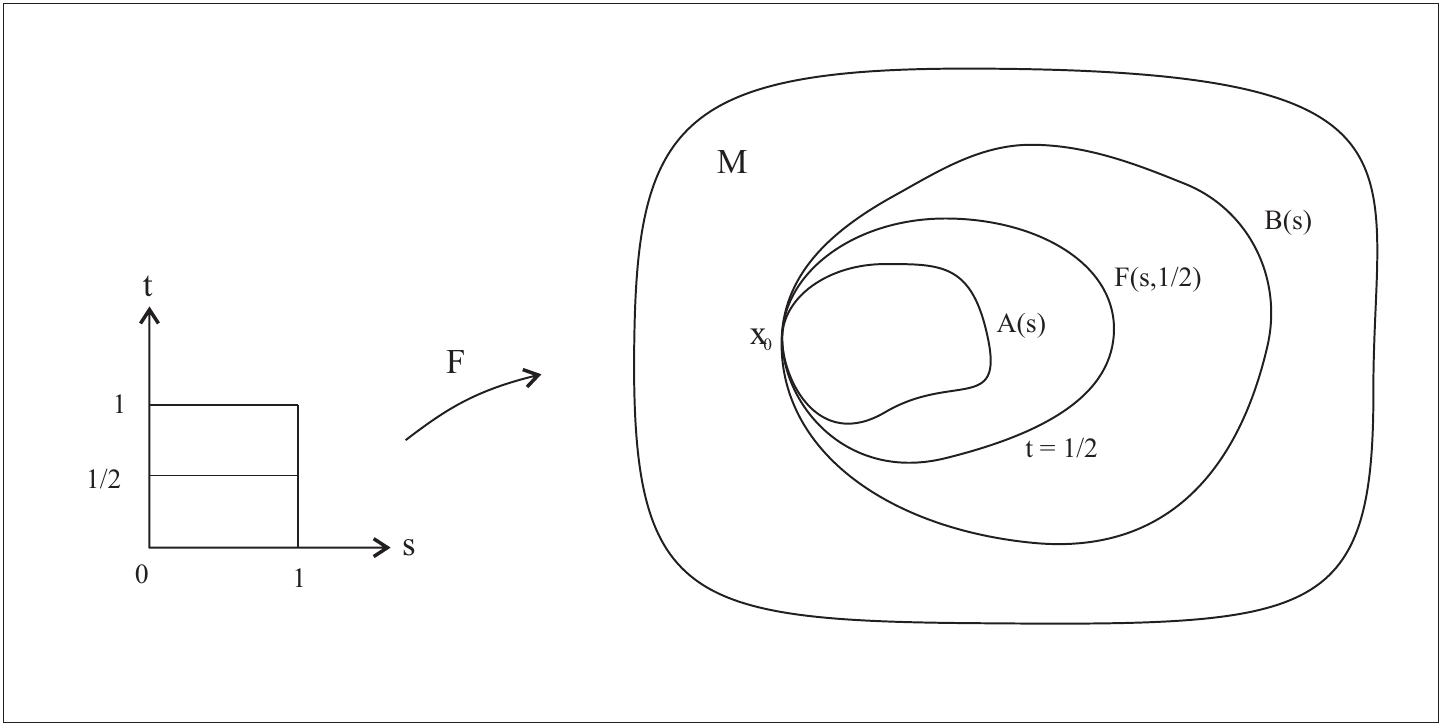}
\end{center}
\caption{In the picture, we can see the representation of the homotopy between
two loops.}%
\label{FigHomotopia}%
\end{figure}

Consider two closed curves $a$ and $b$ with the same base-point $x_{0}$
described by the functions $A(s)$\ and $B(s)$ of the parameter $s$ defined in
the interval $[0,1]$. The loop $a$ is the homotopic path to $b$ $(a\approx b)$ if
there is a continuous function of two parameters $F\left(  s,t\right)  $, with
$t\in \left[0,1\right]$, deforming the loop $a$ into the loop $b$, i.e.,
\begin{align*}
F\left(  s,0\right)   &  =A\left(  s\right) \, ;~\ \ \ \ F\left(  s,1\right)
=B\left(  s\right) \, ;\\
F\left(  0,t\right)   &  =F\left(  1,t\right)  =x_{0} \, .
\end{align*}
Then, $F$ is an homotopy by path between $a$ and $b$. Fig.~\ref{FigHomotopia} illustrates the definition of homotopy.


In addition, homotopy by path is an \emph{equivalence class}:

(i) Trivially, $a\approx a$\ since $F\left(  s,0\right)  = A\left(  s\right)$ is a homotopy. This is, by the way, the identity function;

(ii) Given homotopy $F$ between $a\approx b$, then $G\left(  s,t\right)
=F\left(  s,1-t\right)  $ is homotopic between $b\approx a$;

(iii) If $F$ is a homotopy $a\approx b$ and $G$ is a homotopy $b\approx c$, there
is also a homotopy $H$: $a\approx c$, defined by,%
\[
H\left(  s,t\right)  =\left\{
\begin{array}
[c]{c}%
F\left(  s,2t\right)  ~\ \ \ \ \ \ \ \ \ \ \ \text{for}~\ \ \ t\in
\lbrack0,1/2]\\
G\left(  s,2t-1\right)  ~\ \ \ \ \ \text{for}~\ \ \ t\in\lbrack1/2,1]
\end{array}
\right. \, .
\]

The equivalence class of the loops with a base-point $x_{0}$ satisfy all the axioms of a group. It is the so-called \textbf{fundamental group} or the first homotopy group in $x_{0}$; it is denoted by $\pi_{1}\left(  x_{0}\right)$.

We can parallel transport a vector along a particular loop to be able to measure the curvature of the region inside the loop. When we consider a continuous manifold, it is usual to use an infinitesimal loop, where we can define the \emph{local} curvature, that is, curvature at a point. However, things are a bit different when we deal with a discretized manifold. In this case the curvature is associated to the deficit angle, the loop must include one or more vertexes, and the curvature measurement becomes non-local. Any loop which includes a certain vertex and it is smaller than the perimeter of a fundamental simplex always provides the same value for the deficit angle. Consequently, in a discretized
manifold, the curvature is not a property associated to the infinitesimal loop itself; it is rather a property related to the homotopy group based on the vertex $m$.

In order to parallel transport a vector we need to define a \emph{connection}, that
is, the transport symmetry generator. In our case, the symmetry is the
rotation of the vector $\mathbf{A}$ around the joint $p$. Accordingly, the generator may be taken as the rotation angle $\vec{\sigma}=\sigma~\mathbf{U}_{p}=N~\varepsilon_{p}\mathbf{U}_{p}$ of the vector
$\mathbf{A}$ at the end of the transport. If the transport of $\mathbf{A}$
encompasses several vertexes, it is necessary to add the contribution from all
rotation angles, in a way that the total rotation is given by
\begin{equation}
h=\exp\left(  N~%
{\displaystyle\sum_{p}}
\varepsilon_{p}\mathbf{U}_{p}\right) \, , \label{H}%
\end{equation}
where $N$ is considered as approximately independent of a particular $p$-joint
$(\varrho_{p}\simeq\varrho$), i.e., we take the average of the number of bones
in the manifold's $P$ joints. We call $h\mathbf{A}$ the \textbf{holonomy} of the vector
$\mathbf{A}$ around the vertex $m$. 

Holonomy is closely related to homotopy. In fact, the loops along which we parallel transport vector $\mathbf{A}$ are those in homotopy's definition. Holonomy requires more structure, though: It demands a connection. Even with this additional element, holonomy is also an \emph{equivalence class} since various distinct loops around the same vertex lead to the same value for the deficit angle. There is then an \textbf{holonomy group} at the loops' base-point\footnote{It is noteworthy that $h$ is an operator; it acts upon  $\mathbf{A}$ generating its finite rotation around a given vertex. This fact implies that the set of all possible holonomies around the vertexes of a simplex manifold form a group of transformations \cite{Gilmore}.}. This concept help us to understand the above equation for $h$: A group element can be written as an exponential of the infinitesimal generators of the transformation, just like we see in Eq.~(\ref{H}).

The transport of vector $\mathbf{A}$ along the simplex manifold is done
along the equivalence class loops. Since those loops can be deformed in the
identity loop with base-point $x_{0}$ \cite{Re61}, then the holonomy is the
unity:
\begin{equation}
h=1 \, . \label{H unidade}%
\end{equation}
Indeed, Regge advocates that two arbitrary points $P$ and $Q$ in
contiguous simplexes $T^{m}$ and $T^{m+1}$ can be linked by a curve
$t_{m}$, and that $t_{m}t_{m}^{-1}=u$ is the identity loop with base-point
$P$ related to the joint between $T^{m}$ e $T^{m+1}$. Let's say then that by
repeating this process for all polyhedrons in the trajectory, we define the
composition of curves starting at $P$ and coming back to it. This produces a loop $t_{1}t_{2}t_{3}...t_{n}$ homotopic to the identity loop $u$,%
\[
t_{1}t_{2}t_{3}...t_{n}\approx u \, .
\]
($n$ is the number of polyhedrons in the transport. Note that a cyclic order of $1$ to $n$ was defined; this is arbitrary, but general). We arrive in the unitary holonomy in the Eq. (\ref{H unidade}) by using the identity loop as an equivalence class representative.

Substituting (\ref{H unidade})\ into (\ref{H}):
\[
\exp\left(  N~%
{\displaystyle\sum_{p}}
\varepsilon_{p}\mathbf{U}_{p}\right)  =1 \, .
\]
Therefore, the exponential's argument must vanish:%
\[%
{\displaystyle\sum_{p}}
\varepsilon_{p}\mathbf{U}_{p}=0 \, .
\]
By dualization, this is the same as:
\begin{equation}%
{\displaystyle\sum_{p}}
\varepsilon_{p}U_{\mu\nu}^{p}=0 \, . \label{simetria homotopia}%
\end{equation}
This result followed from topological considerations. It will be key to derive Bianchi's identity associated to the curvature on discretized manifold.


\subsubsection{Relating $\varrho_{p}$ to $\varrho$ \label{sec-rho}}

The local density of bones in the $p$-th joint, $\varrho_{p}$, cannot be
constant throughout the manifold because there are regions on the manifold with higher curvature and consequently with higher concentration of polyhedrons. The rate of variation of $\varrho_{p}$ can be obtained as a function of the average bones density $\varrho$.

\begin{figure}[h]
\begin{center}
\includegraphics[scale=0.6]{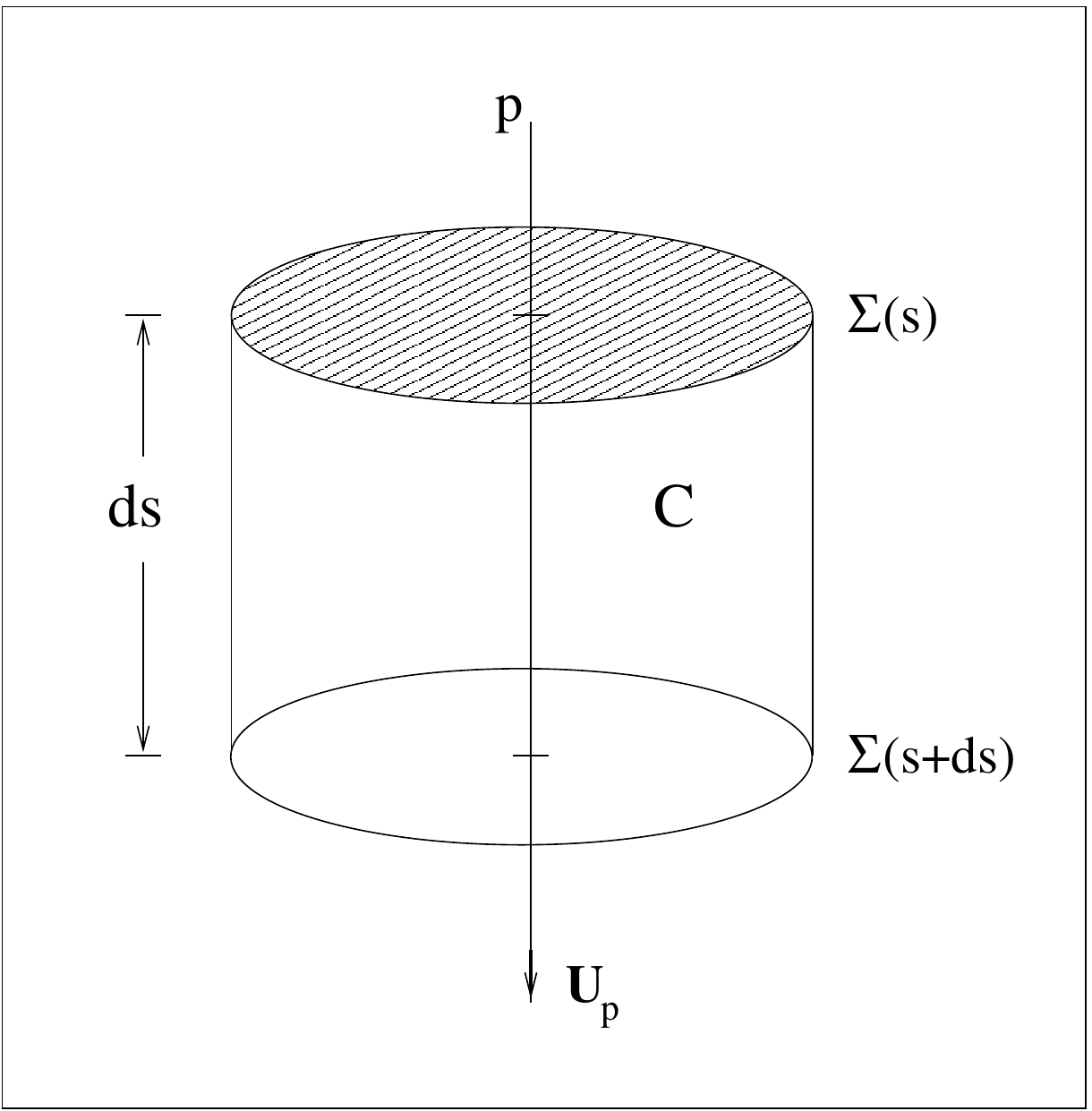}
\end{center}
\caption{Infinitesimal cylinder $C$ around the sheaf $p$ of edges.}%
\label{FigCilindro}%
\end{figure}

The density $\varrho_{p}$ is the number of bones piercing through the surface
$\Sigma$ orthogonal to $\mathbf{U}^{p}$\ ($\varrho_{p}$ is a superficial
density). We refer the reader to Fig.~\ref{FigCilindro}. Let $s$ be a parameter along the bundle of bones $p$ with origin at the vertex, i.e., $s$ is along the direction of $\mathbf{U}^{p}$. The position of $\Sigma$
is determined by $s$. Now, let $C$ be a cylinder of base equals $\Sigma$\ and hight $ds$.


The number of bones inside the cylinder is the number of bones leaving $C$,
\[
\varrho_{p}\left(  s+ds\right)  ~\Sigma \, ,
\]
minus the number of bones entering $C$,
\[
\varrho_{p}(s)~\Sigma \, .
\]
On the other hand, the number of bones within $C$ must be the average density
$\varrho$ times the volume $\Sigma~ds$ of the cylinder:
\[
\varrho\left(  s\right)  ~\Sigma~ds \, .
\]
($\varrho$ is a volumetric density). Hence,
\[
\varrho_{p}\left(  s+ds\right)  ~\Sigma-\varrho_{p}(s)~\Sigma=\varrho\left(
s\right)  ~\Sigma~ds \, ,
\]
i.e.,
\begin{equation}
\varrho=\frac{d\varrho_{p}}{ds} \, . \label{rho direcional}
\end{equation}
The derivative in (\ref{rho direcional}) is the directional derivative along
the bundle of bones, that is, along the  $\mathbf{U}^{p}$ direction.
In the Cartesian coordinate system, this equation reads:
\begin{equation}
\varrho=\frac{dx^{\mu}}{ds}\frac{d\varrho_{p}}{dx^{\mu}}=U_{p}^{\mu}%
\frac{d\varrho_{p}}{dx^{\mu}}=\mathbf{U}_{p}\cdot\mathbf{\nabla}\varrho
_{p} \, .\label{rho grad}%
\end{equation}
Notice the identification $U_{p}^{\mu}=dx^{\mu}/ds$ is reasonable because $dx^{\mu}/ds$
is the ``velocity vector'' \emph{along} the bone.


\subsubsection{Calculating $B_{\lambda\alpha\beta\mu\nu}$\ in the discretized
manifold}

Keeping in mind the results in Sections \ref{sec-homotopia} and \ref{sec-rho}, let us return to the problem of finding the discrete version of the second Bianchi identity.

According to (\ref{Riemann simpletico}):
\begin{equation}
R_{\alpha\beta\mu\nu}=\sum\limits_{p}\varrho_{p}\varepsilon_{p}U_{~\alpha
\beta}^{p}U_{~\mu\nu}^{p} \, . \label{Riemann somado}%
\end{equation}
The sum was included because the parallel transport of $\mathbf{A}$ eventually encloses many joints in the simplex manifold.

In our approximation to discretize the space into Euclidian polyhedrons, the
covariant derivatives in the definition of $B_{\lambda\alpha\beta\mu\nu}$,
\[
B_{\lambda\alpha\beta\mu\nu}\equiv\nabla_{\lambda}R_{\alpha\beta\mu\nu}%
+\nabla_{\alpha}R_{\beta\lambda\mu\nu}+\nabla_{\beta}R_{\lambda\alpha\mu\nu} \, ,
\]
reduce to ordinary derivatives (since there is no curvature in the polyhedrons):\footnote{The $\Gamma$ cancel out throughout the whole manifold, not just locally --- see Eq.~(\ref{derivada covariante}).}%
\begin{equation}
B_{\lambda\alpha\beta\mu\nu}\equiv\partial_{\lambda}R_{\alpha\beta\mu\nu
}+\partial_{\alpha}R_{\beta\lambda\mu\nu}+\partial_{\beta}R_{\lambda\alpha
\mu\nu} \, . \label{B}%
\end{equation}
Next, we show Eq. (\ref{B}) vanishes, indeed.

The first step is to introduce (\ref{Riemann somado}) in (\ref{B}) and take the derivatives there indicated. $\mathbf{U}^{p}$\ is constant. The deficit angle in the $p$-joint is constant:\footnote{See also Section \ref{sec-delta-deficit}.}
\[
\partial_{\mu}\varepsilon_{p}=0 \, .
\]
Therefore, 
\begin{equation}
B_{\lambda\alpha\beta\mu\nu}=\sum\limits_{p}\varepsilon_{p}U_{~\mu\nu}%
^{p}\left(  U_{~\alpha\beta}^{p}\partial_{\lambda}\varrho_{p}+U_{~\beta
\lambda}^{p}\partial_{\alpha}\varrho_{p}+U_{~\lambda\alpha}^{p}\partial
_{\beta}\varrho_{p}\right) \, . \label{B(U)}%
\end{equation}
Now, we make use of the identity\footnote{Identity (\ref{identidade}) will be demonstrated at the end of this section, below Eq.~(\ref{Bianchi simpletica}).}
\begin{equation}
\left(  U_{~\alpha\beta}^{p}\partial_{\lambda}+U_{~\beta\lambda}^{p}%
\partial_{\alpha}+U_{~\lambda\alpha}^{p}\partial_{\beta}\right)
=\epsilon_{\alpha\beta\lambda}~\mathbf{U}_{p}\cdot\mathbf{\nabla}
\label{identidade}%
\end{equation}
to write (\ref{B(U)}) as:%
\[
B_{\lambda\alpha\beta\mu\nu}=\sum\limits_{p}\varepsilon_{p}U_{~\mu\nu}%
^{p}\epsilon_{\alpha\beta\lambda}~\mathbf{U}_{p}\cdot\mathbf{\nabla}%
\varrho_{p} \, .
\]
Next, by substituting the result (\ref{rho grad}) from the last section into this
equation, we have:%
\[
B_{\lambda\alpha\beta\mu\nu}=\epsilon_{\alpha\beta\lambda}~\varrho
\sum\limits_{p}\varepsilon_{p}U_{\mu\nu}^{p} \, .
\]
Finally, Eq.~(\ref{simetria homotopia}) from Section \ref{sec-homotopia}, lead us to:
\begin{equation}
B_{\lambda\alpha\beta\mu\nu}=0 \, .\label{Bianchi simpletica}%
\end{equation}
This is the second Bianchi identity for the discretized version of gravitation.

In order to feel completely satisfied with the demonstration of (\ref{Bianchi simpletica}), we should also derive (\ref{identidade}). In fact,
\begin{align*}
\mathbf{U}_{p}\cdot\mathbf{\nabla}  &  =U_{p}^{\mu}\partial_{\mu}=\frac{1}%
{2}\epsilon^{\alpha\beta\mu}U_{~\alpha\beta}^{p}\partial_{\mu}=\\
&  =\frac{1}{2}\left[  \frac{1}{3}\left(  \epsilon^{\alpha\beta\mu}%
+\epsilon^{\beta\mu\alpha}+\epsilon^{\mu\alpha\beta}\right)  \right]
U_{~\alpha\beta}^{p}\partial_{\mu}
\end{align*}
by the cyclic property of $\epsilon^{\alpha\beta\mu}$ indexes. Distributing
$U_{~\alpha\beta}^{p}\partial_{\mu}$ and renaming dummy indexes,
\[
\mathbf{U}_{p}\cdot\mathbf{\nabla}=\frac{1}{6}\epsilon^{\alpha\beta\mu}\left[
U_{~\alpha\beta}^{p}\partial_{\mu}+U_{~\mu\alpha}^{p}\partial_{\beta
}+U_{~\beta\mu}^{p}\partial_{\alpha}\right]  \, .
\]
Hence:%
\[
\epsilon_{\alpha\beta\mu}~\mathbf{U}_{p}\cdot\mathbf{\nabla}=\frac{1}%
{6}\left(  \epsilon_{\alpha\beta\mu}\epsilon^{\alpha\beta\mu}\right)  \left[
U_{~\alpha\beta}^{p}\partial_{\mu}+U_{~\mu\alpha}^{p}\partial_{\beta
}+U_{~\beta\mu}^{p}\partial_{\alpha}\right]  \, .
\]
From Eq.~(\ref{Levi contracao}):%
\[
\left(  \epsilon_{\alpha\beta\mu}\epsilon^{\alpha\beta\mu}\right)  =\left(
\delta_{~\alpha}^{\alpha}\delta_{~\beta}^{\beta}-\delta_{~\beta}^{\alpha
}\delta_{~\alpha}^{\beta}\right)  =\left(  3^{2}-\delta_{~\alpha}^{\alpha
}\right)  =9-3=6 \, .
\]
Substituting this in the above expression results in:%
\[
\epsilon_{\alpha\beta\mu}~\mathbf{U}_{p}\cdot\mathbf{\nabla}=U_{~\alpha\beta
}^{p}\partial_{\mu}+U_{~\mu\alpha}^{p}\partial_{\beta}+U_{~\beta\mu}%
^{p}\partial_{\alpha} \, ,
\]
which is exactly the identity (\ref{identidade}) we wanted demonstrated.

We are now ready to build the skeleton version fo Einstein field equation for gravity. This will be accomplished in the upcoming section.


\section{Action Integral and Einstein Equations in Simplex
Gravity \label{sec-EOM}}


\subsection{Regge's action}

The sourceless gravity field equations are found by varying Einstein-Hilbert action \cite{Sabbata}%
\begin{equation}
I=\frac{1}{16\pi}\int d^{4}x~\sqrt{-g}~R\label{acao RG}%
\end{equation}
with respect to the metric tensor $g^{\mu\nu}$. This tensor encapsulates all information about curvature: Christoffel symbols and the curvature tensor are written in terms of the metric and its derivatives. In fact, 
\begin{equation}
\frac{\delta I}{\delta g^{\mu\nu}}=0 \, , \label{delta I RG}%
\end{equation}
leads to \cite{Mi17}:
\begin{equation}
R_{\mu\nu}-\frac{1}{2}g_{\mu\nu}R=0 \, . \label{Einstein RG}%
\end{equation}

Our next task is to suggest a version of the equations above for the simplex manifold. The integral in (\ref{acao RG}) is translated to a sum over all the $p$ joints. In Section \ref{sec-curvatura} we obtained the discretized version of the scalar curvature: $R=2\varrho_{p}\varepsilon_{p}$, cf. Eq.~(\ref{R}). What is the analogous of the measure $d^{4}x~\sqrt{-g}$? Regge suggests that the adequate measure for the discretized action $I$ is the quantity $L_{p}= L_p \left(\varrho_p \right)$ defined at the $p$-joint. In three dimensions $L_p$ is simply the length $l_{p}$ of the bones in the hinge. In four dimensions $L_{p}$ is actually the area of the simplexes edges at the $p$-joint. Thus,
\[
\frac{1}{16\pi}\int d^{4}x~\sqrt{-g}~R\rightarrow\frac{1}{16\pi}%
\sum\limits_{p} L_{p} \left(  2 \varepsilon_{p} \right) \, ,
\]
where the density $\varrho_{p}$ coming from $R$ was included into the functional form of $L_p=L_p(\varrho_{p})$. The action according to Regge is, then:
\begin{equation}
I=\frac{1}{8\pi}\sum\limits_{p}\varepsilon_{p}L_{p} \, . \label{acao simpletica}%
\end{equation}

The measure $L_{p}$ can be written in terms of the length $l_{p}$ of the
$p$-joint (according to Section \ref{sec-area} below). We also know that the
length of the bones yields the same type of information about the skeleton manifold that the metric
provides for the continuous manifold. For this reason, we choose to take $l_{p}$ as the variation parameter of $I$. Accordingly, the analogous of (\ref{delta I RG}) is:
\begin{equation}
\frac{\delta I}{\delta l_{p}}=0 \, ,\label{delta I simpletica}%
\end{equation}
which, in view of Eq.~(\ref{acao simpletica}), reads:
\begin{equation}
\delta I=\frac{1}{8\pi}\sum\limits_{p}\delta\varepsilon_{p}L_{p}+\frac{1}%
{8\pi}\sum\limits_{p}\varepsilon_{p}\delta L_{p} \, . \label{var I}%
\end{equation}

We need another long digression (Section \ref{sec-delta-deficit}) to show that
the first term on the right-hand side of (\ref{var I}) is zero. This is rather surprising since
$\varepsilon_{p}$ depends on $l_{p}$. However, the effort is necessary to derive the discrete version of Einstein equations. In Section \ref{sec-area} we find $\delta L_{p}$ as a function of $l_{p}$. Finally, Regge field equations are built in Section \ref{sec-Einstein}.


\subsection{Checking $\sum\limits_{p}\delta\varepsilon_{p}L_{p}=0$
\label{sec-delta-deficit}}

\begin{figure}[H]
\begin{center}
\includegraphics[scale=0.6]{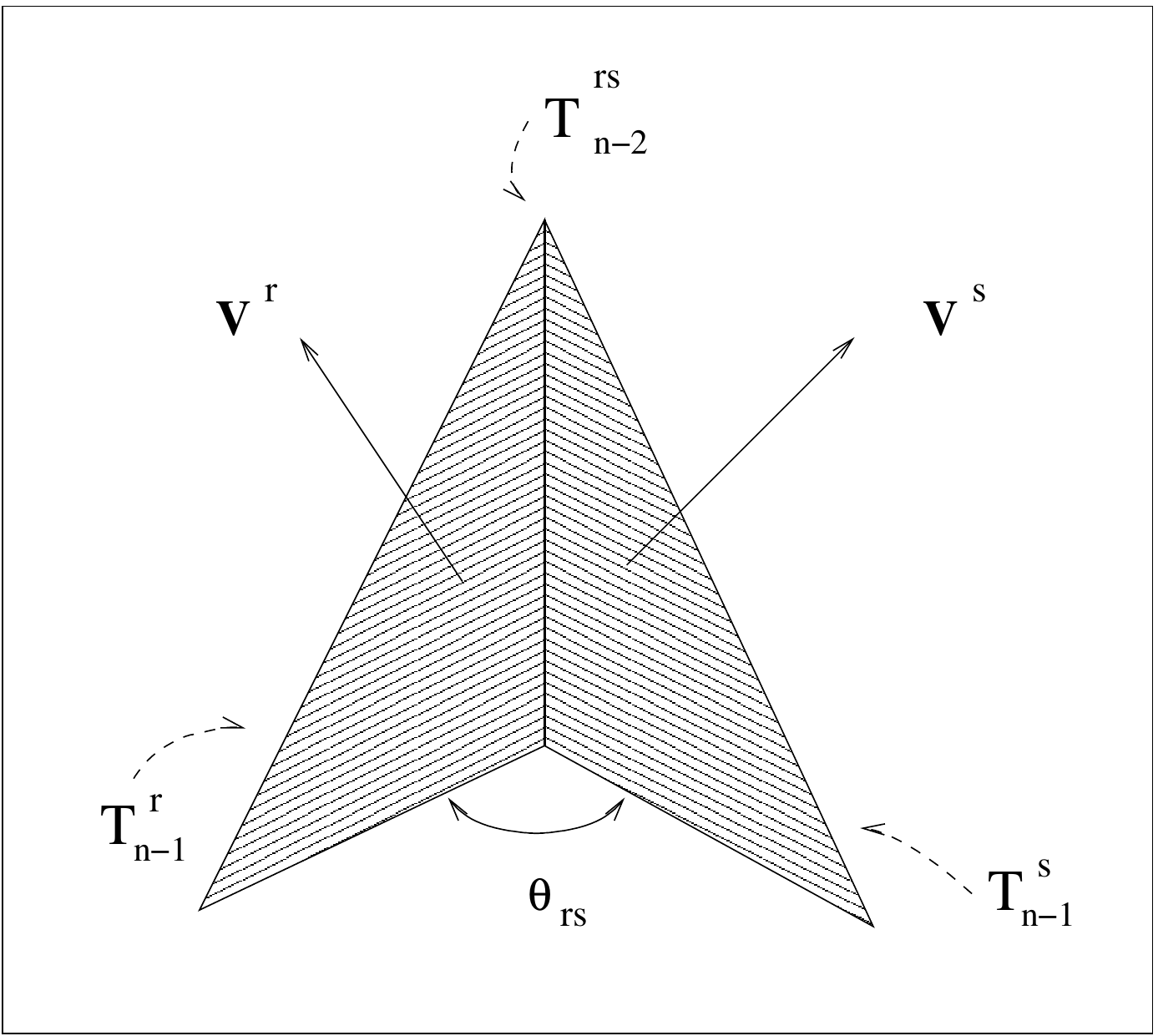}
\end{center}
\caption{Two-dimensional representation of the simplexes $T_{n-1}^{r}$ and
$T_{n-1}^{s}$\ and their comum edge $T_{n-2}^{rs}$. The unit vectors $V_{\mu
}^{r}\ $and $V_{\nu}^{s}$ are normal to $T_{n-1}^{r}$ and $T_{n-1}^{s}$.}
\label{FigSimplexos}
\end{figure}

Consider the following example: a triangle is a two-dimensional simplex represented by
$T_{2}$. The edges of $T_{2}$ are three straight segments, which are the unidimensional simplexes, $T_{1}$. In order to assemble $T_{2}$, it was necessary to gather $2+1=3$ simplexes $T_{1}$. %
The fundamental simplex in three dimensions is the tetrahedron. A tetrahedron $T_{3}$ is
composed of $3+1=4$ triangles $T_{2}$. This can be generalized: Let
$T_{n}$ be a simplex of dimension $n$. The edges are simplexes $T_{n-1}$. It
is necessary $\left(  n+1\right)  $ simplexes $T_{n-1}$ to generate a
fundamental $T_{n}$ simplex, eventually used to discretize the manifold.
Let $r$ and $s$ be the labels used to identify the edge-simplexes $T_{n-1}$,
so that $r,s=0,1,2,...,\left(  n+1\right) $. Fig.~\ref{FigSimplexos} shows a two-dimensional representation of $T_{n-1}^{r}$.


Any two contiguous simplexes $T_{n-1}^{r}$\ and $T_{n-1}^{s}$ share a common
edge. According to the previous discussion, the edge has the dimension $\left(
n-2\right)  $ and label $rs$. The edge is, then, denoted by $T_{n-2}^{rs}$.
By definition, $\theta_{rs}$\ is the angle in between $T_{n-1}^{r}$ and $T_{n-1}^{s}$.

Now we define unitary vectors $V_{\mu}^{r}$\ and $V_{\mu}^{s}$ normal to (the \textquotedblleft surface\textquotedblright\ of) $T_{n-1}^{r}$ and $T_{n-1}^{s}$. The index $\mu$\ refers to the components of
$\mathbf{V}$\ in a Cartesian coordinate system defined in the manifold.

The following identities hold:

\begin{enumerate}
\item Unitary norm:%
\begin{equation}
\left\Vert \mathbf{V}^{r}\right\Vert ^{2}=V_{~\mu}^{r}V_{r}^{~\mu}=1 \, .
\label{norma V}%
\end{equation}
(The position of the label $r$ is irrelevant; but the position is important for $\mu$, which
respects Einstein's summation convention). The same is valid for $\mathbf{V}^{s}$.

\item Dot product:%
\[
\mathbf{V}^{r}\cdot\mathbf{V}^{s}=\left\Vert \mathbf{V}^{r}\right\Vert
\left\Vert \mathbf{V}^{s}\right\Vert \cos\theta_{rs} \, ,%
\]
i.e.,%
\begin{equation}
V_{~\mu}^{r}V^{s\mu}=\cos\theta_{rs} \, . \label{theta}%
\end{equation}

\end{enumerate}

Consider the antisymmetric tensor $\mathbf{V}^{rs}$,%
\begin{equation}
V_{\mu\nu}^{rs}=-V_{\nu\mu}^{rs} \, , \label{antissimetria}%
\end{equation}
inspired by the vector product definition:
\begin{equation}
V_{\mu\nu}^{rs}=\frac{1}{\sin\theta_{rs}}\left(  V_{~\mu}^{r}V_{~\nu}%
^{s}-V_{~\nu}^{r}V_{~\mu}^{s}\right)  \, , \label{V mu nu}%
\end{equation}
where the factor $\left(  1/\sin\theta_{rs}\right)  $\ may be seen as a
normalization factor\footnote{We can understand this normalization factor by
recalling:
\begin{align*}
\mathbf{x}\times\mathbf{y}  &  =\left\vert
\begin{pmatrix}
\mathbf{i} & \mathbf{j} & \mathbf{k}\\
x_{1} & x_{2} & x_{3}\\
y_{1} & y_{2} & y_{3}%
\end{pmatrix}
\right\vert =\\
&  =\left(  x_{2}y_{3}-x_{3}y_{2},x_{3}y_{1}-x_{1}y_{3},x_{2}y_{3}-x_{3}%
y_{2}\right) \, ,
\end{align*}
i.e.,%
\[
\left(  \mathbf{x}\times\mathbf{y}\right)  _{i}=\sum_{j,k}\epsilon
_{ijk}\left(  x_{j}y_{k}-y_{k}x_{j}\right)
\]
and%
\[
\left\Vert \mathbf{x}\times\mathbf{y}\right\Vert =\left\Vert \mathbf{x}%
\right\Vert \left\Vert \mathbf{y}\right\Vert \sin\theta_{xy} \, .
\]
If (as in our case), $\left\Vert \mathbf{x} \right\Vert = \left\Vert \mathbf{y}\right\Vert =1$, then
\[
\frac{\mathbf{x}\times\mathbf{y}}{\left\Vert \mathbf{x}\times\mathbf{y}%
\right\Vert }=\frac{1}{\sin\theta_{xy}}\sum_{j,k}\epsilon_{ijk}\left(
x_{j}y_{k}-y_{k}x_{j}\right)  \, ,
\]
which is analogous to (\ref{V mu nu}).} of $V_{\mu\nu}^{rs}$.

\begin{figure}[h]
\begin{center}
\includegraphics[scale=0.6]{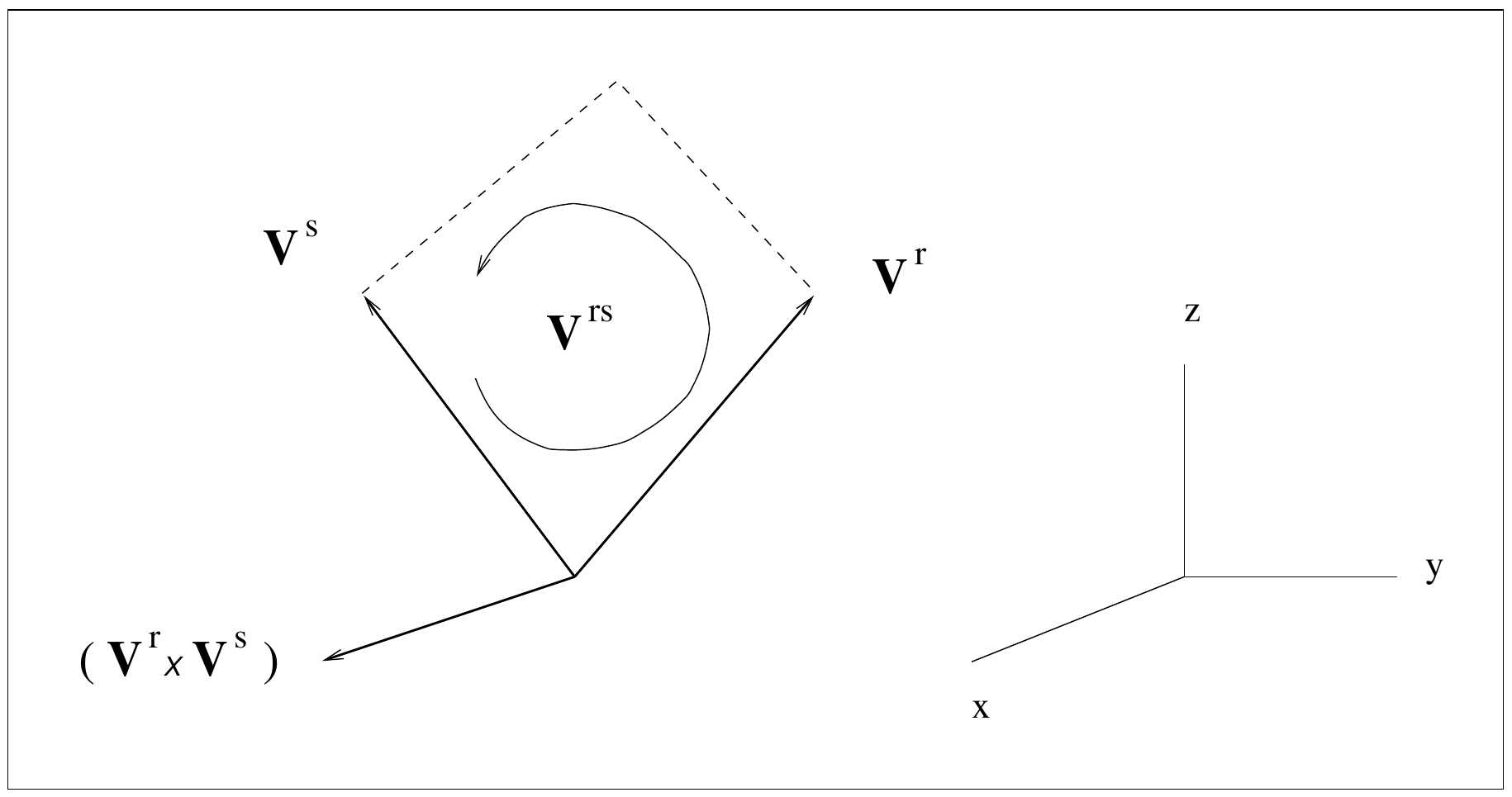}
\end{center}
\caption{Pictorically, $\left(  \mathbf{V}^{r}\times\mathbf{V}^{s}\right)
$\ is ``orthogonal'' to $\mathbf{V}^{rs}$.}%
\label{FigOrtogonais}%
\end{figure}

However, there is a fundamental difference between the definition $\mathbf{V}^{rs}$ 
and the cross product $\left(  \mathbf{V}^{r}\times\mathbf{V}^{s}\right)  $, namely the 
\textbf{dualization} process. While $\left(  \mathbf{V}^{r}\times\mathbf{V}%
^{s}\right)  $ is an axial vector, $\mathbf{V}^{rs}$ is its dual quantity. That
means $\left(  \mathbf{V}^{r}\times\mathbf{V}^{s}\right)  $ is orthogonal
to $\mathbf{V}^{r}$\ and to $\mathbf{V}^{s}$, i.e. it is orthogonal to the
area defined by them -- see Fig.~\ref{FigOrtogonais}. On the other hand, $\mathbf{V}^{rs}$ can be understood
as a quantity over the area, defining an orientation thereof.


The contraction $V_{\mu\nu}^{rs}V_{rs}^{\mu\nu}$ of
the antisymmetric tensors is the following:\footnote{We emphasize that there is no implied sum in the repeated indexes $r$ and $s$. These indices only identify normal unit vectors to the surfaces of contiguous simplexes. On the other hand, the Greek indexes, such as $\mu$ and $\nu$, refer to Cartesian coordinates and are subject to the Einstein's sum convention.}
\begin{align*}
V_{\mu\nu}^{rs}V_{rs}^{\mu\nu}  &  =\frac{1}{\sin^{2}\theta_{rs}}\left[
\left(  V_{\mu}^{r}V_{r}^{\mu}\right)  \left(  V_{\nu}^{s}V_{s}^{\nu}\right)
-\left(  V_{\mu}^{r}V_{r}^{\nu}\right)  \left(  V_{\nu}^{s}V_{s}^{\mu}\right)
+\right. \\
&  \left.  -\left(  V_{\nu}^{r}V_{r}^{\mu}\right)  \left(  V_{\mu}^{s}%
V_{s}^{\nu}\right)  +\left(  V_{\nu}^{r}V_{r}^{\nu}\right)  \left(  V_{\mu
}^{s}V_{s}^{\mu}\right)  \right]  \, .
\end{align*}
From (\ref{norma V}) and (\ref{theta}):
\[
V_{\mu\nu}^{rs}V_{rs}^{\mu\nu}=\frac{1}{\sin^{2}\theta_{rs}}\left[
2-2\cos^{2}\theta_{rs}\right]  \, ,
\]
that is,%
\begin{equation}
V_{\mu\nu}^{rs}V_{rs}^{\mu\nu}=2~. \label{norma V mu nu}%
\end{equation}

Now, take the variation of (\ref{theta}):%
\[
\delta\left(  V_{~\mu}^{r}V^{s\mu}\right)  =\delta\left(  \cos\theta
_{rs}\right) \, ,
\]
or%
\[
\delta V_{~\mu}^{r}V^{s\mu}+V_{~\mu}^{r}\delta V^{s\mu}=-\sin\theta
_{rs}~\delta\theta_{rs} \, ,
\]
or yet,%
\begin{equation}
\delta\theta_{rs}=-\frac{1}{\sin\theta_{rs}}\left(  V_{~\mu}^{r}\delta
V^{s\mu}+V_{~\mu}^{s}\delta V^{r\mu}\right)  \, . \label{delta theta}%
\end{equation}

It will be handy to rewrite this equation in terms of the object $V_{\mu\nu
}^{rs}$. For this end, we contract $V_{\mu\nu}^{rs}$, Eq. (\ref{V mu nu}), with
$V_{r}^{\mu}$:
\begin{align*}
V_{\mu\nu}^{rs}V_{r}^{\mu}  &  =\frac{1}{\sin\theta_{rs}}\left(  V_{~\mu}%
^{r}V_{r}^{\mu}V_{~\nu}^{s}-V_{~\nu}^{r}V_{r}^{\mu}V_{~\mu}^{s}\right)  =\\
&  =\frac{1}{\sin\theta_{rs}}\left(  V_{~\nu}^{s}-V_{~\nu}^{r}\cos\theta
_{rs}\right)  \, ,
\end{align*}
where we have used (\ref{norma V}) and (\ref{theta}) again. Next, consider the
contraction with $\delta V_r^{\nu}$:%
\begin{equation}
V_{\mu\nu}^{rs}V_{r}^{\mu}\delta V_{r}^{\nu}=\frac{1}{\sin\theta_{rs}}\left(
V_{~\nu}^{s}\delta V_{r}^{\nu}-\cos\theta_{rs}V_{~\nu}^{r}\delta V_{r}^{\nu
}\right)  \, . \label{VVdeltaV}
\end{equation}
However, the last term cancels out. This follows from (\ref{norma V}):
\[
\delta\left(  V_{~\mu}^{r}V_{r}^{~\mu}\right)  =\delta\left(  1\right)  =0 \, ,
\]
i.e.,%
\[
\delta V_{~\mu}^{r}V_{r}^{~\mu}+V_{~\mu}^{r}\delta V_{r}^{~\mu}=0~.
\]
Raising and lowering the indexes in the first term, it results:
\[
\delta V_{~\mu}^{r}V_{r}^{~\mu}=0 \, ,
\]
as stated. Therefore, identity (\ref{VVdeltaV}) reduces to:
\[
V_{\mu\nu}^{rs}V_{r}^{\mu}\delta V_{r}^{\nu}=\frac{1}{\sin\theta_{rs}}V_{~\nu
}^{s}\delta V_{r}^{\nu} \, ,
\]
or,
\begin{equation}
V_{~\nu}^{s}\delta V_{r}^{\nu}=\sin\theta_{rs}\left(  V_{\mu\nu}^{rs}%
V_{r}^{\mu}\delta V_{r}^{\nu}\right)  \, . \label{V var V 1}%
\end{equation}

This is the second term of the expression (\ref{delta theta}) for
$\delta\theta_{rs}$. The first term, $V_{~\mu}^{r}\delta V^{s\mu}$, comes from
an entirely analogous process:%
\begin{align*}
V_{\mu\nu}^{rs}V_{s}^{\mu} &  =\frac{1}{\sin\theta_{rs}}\left(  V_{~\mu}%
^{r}V_{s}^{\mu}V_{~\nu}^{s}-V_{~\nu}^{r}V_{~\mu}^{s}V_{s}^{\mu}\right)  =\\
&  =\frac{1}{\sin\theta_{rs}}\left(  V_{~\nu}^{s}\cos\theta_{rs}-V_{~\nu}%
^{r}\right) \, .
\end{align*}%
\[
V_{\mu\nu}^{rs}V_{s}^{\mu}\delta V_{s}^{\nu}=\frac{1}{\sin\theta_{rs}}\left(
\cos\theta_{rs}V_{~\nu}^{s}\delta V_{s}^{\nu}-V_{~\nu}^{r}\delta V_{s}^{\nu
}\right)  =-\frac{1}{\sin\theta_{rs}}V_{~\nu}^{r}\delta V_{s}^{\nu} \, .
\]%
\begin{equation}
V_{~\nu}^{r}\delta V^{s\nu}=-\sin\theta_{rs}V_{\mu\nu}^{rs}V_{s}^{\mu}\delta
V_{s}^{\nu} \, .\label{V var V 2}%
\end{equation}

Substituting (\ref{V var V 1}) and (\ref{V var V 2}) into Eq. (\ref{delta theta}):
\begin{equation*}
\delta\theta_{rs}=-\frac{1}{\sin\theta_{rs}}\left(  -\sin\theta_{rs}V_{\mu\nu
}^{rs}V_{s}^{\mu}\delta V_{s}^{\nu}+\sin\theta_{rs}V_{\mu\nu}^{rs}V_{r}^{\mu
}\delta V_{r}^{\nu}\right) \, . 
\end{equation*}
Renaming the repeated indexes in the second term and using the
antisymmetry of $V_{\mu\nu}^{rs}$:
\begin{equation}
\delta\theta_{rs}=-V_{\mu\nu}^{rs}\left(  V_{s}^{\mu}\delta V_{s}^{\nu}%
+V_{r}^{\nu}\delta V_{r}^{\mu}\right) \, . \label{delta theta V mu nu}%
\end{equation}

The deficit angle $\varepsilon_{p}$\ associated to the node $p$ is given in
term of the sum of dihedral angles $\theta_{rs}$ on that node,\footnote{Notice 
the sum is carried over the pair $(rs)$ labeling the edge in between $T_{(n-1)}^{r}$\ and $T_{(n-1)}^{s}$.}
\[
\varepsilon_{p}=2\pi-\sum_{\left(  rs\right)  }\theta_{rs} \, .
\]
The change in $\varepsilon_{p}$\ is, therefore, written as the function of
$\delta\theta_{rs}$:%
\[
\delta\varepsilon_{p}=-\sum_{\left(  rs\right)  }\delta\theta_{rs} \, .
\]
The sum of $\delta\varepsilon_{p}$ over all the $p$-joints weighted by the generalized area  $L_{p}$ of the joint is:
\begin{align}
\sum\limits_{p}\delta\varepsilon_{p}L_{p}  &  =-\sum\limits_{p}\sum_{\left(
rs\right)  }\delta\theta_{rs}L_{rs}^{p} = \nonumber \\
&  =\sum\limits_{p}\sum_{\left(  rs\right)  }\left(  V_{s}^{\mu}\delta
V_{s}^{\nu}+V_{r}^{\nu}\delta V_{r}^{\mu}\right)  V_{\mu\nu}^{rs}L_{rs}^{p} \, , \label{soma deltaE Lp}
\end{align}
where we have used (\ref{delta theta V mu nu}). $L_{rs}^{p}$ is the
contribution for the measure from each joint. 

\begin{figure}[H]
\begin{center}
\includegraphics[scale=0.5]{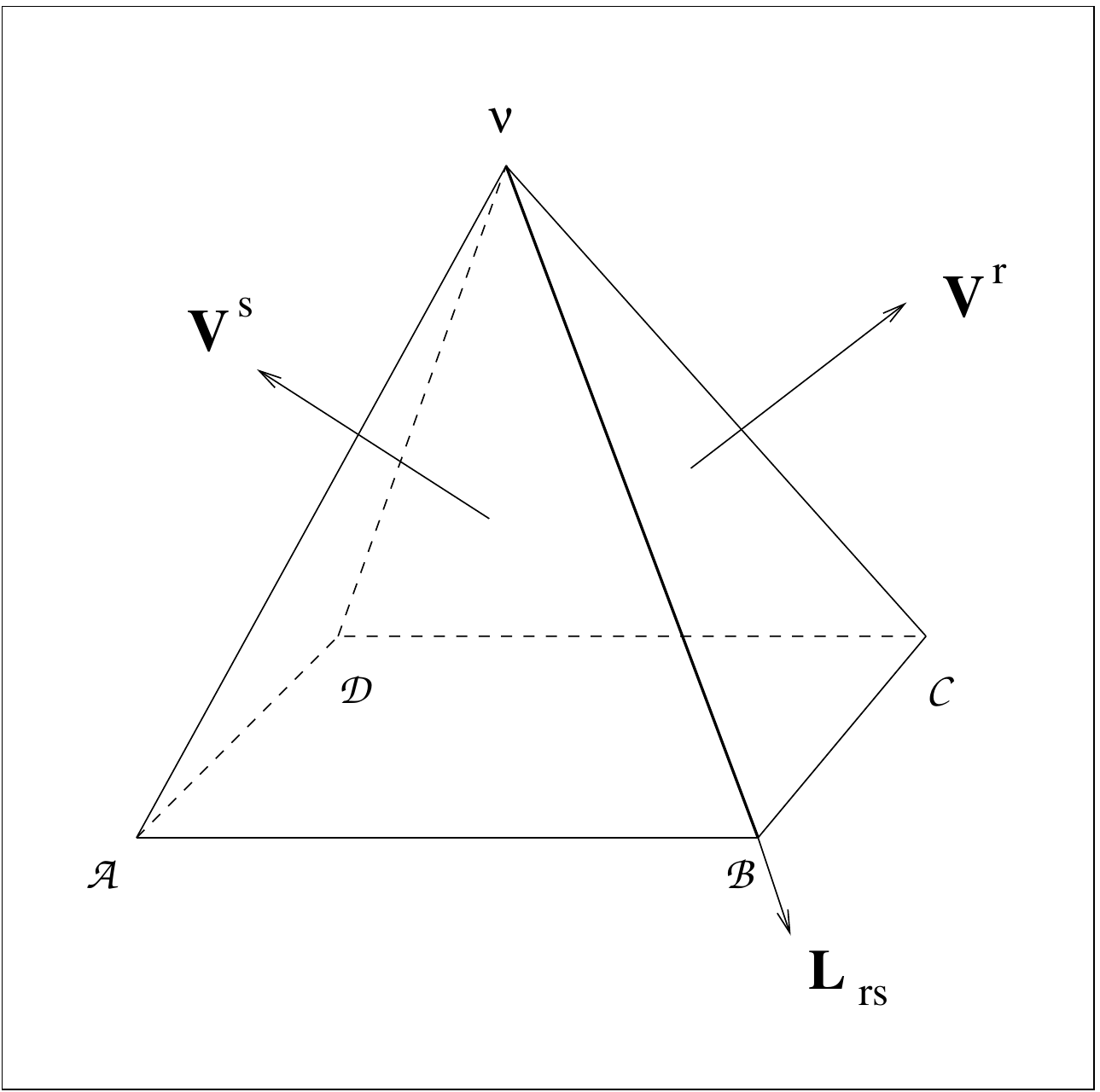}
\end{center}
\caption{The pyramidal simplex example allows us to understand that the sum over the edges $(r,s)$ is equivalent to the sum over the adjoint faces $r$ and $s$. See Eq.~(\ref{quebra soma}) and discussion below it.}
\label{FigPiramide}
\end{figure}

Note that the sum over the pair $\left(  rs\right)  $ labeling a
particular edge is equivalent to a sum over the adjoint surfaces $r$ and $s$:
\begin{equation}
\sum_{\left(  rs\right)  }\rightarrow\sum_{r}\sum_{s}\text{ \ \ or \ \ }%
\sum_{s}\sum_{r} \, . \label{quebra soma}%
\end{equation}
In effect, consider the pyramid vertex $\mathcal{V}$ in Fig.~\ref{FigPiramide}. 
The vertex $\mathcal{V}$ is related to the joint $p$ of interest here. The index $(rs)$ indicates, for example, the edge $\mathcal{VB}$ when we defined $\mathbf{V}^{r}$ on the face $\mathcal{VBC}$ and $\mathbf{V}^{s}$ on the face $\mathcal{VAB}$. 


The sum over $(rs)$ means adding the contributions of all the edges $\mathcal{VB}$, $\mathcal{VC}$,
$\mathcal{VD}$ and $\mathcal{VA}$. For that end, we subsequently place the pair $\mathbf{V}^{r}$ 
and $\mathbf{V}^{s}$ on different pairs of faces: First, $\mathbf{V}^{r}$ is placed on 
the face $\mathcal{VBC}$ and $\mathbf{V}^{s}$ is placed on the face $\mathcal{VAB}$ (Fig.~\ref{FigPiramide}); this counts the contribution by the edge $\mathcal{VB}$. After that, we take $\mathbf{V}^{r}$ on $\mathcal{VAB}$ and $\mathbf{V}^{s}$ on $\mathcal{VDA}$ to account for the contribution of the edge $\mathcal{VA}$; and so on. The act of changing from one edge to the next means to compute one more term in the sum over $(rs)$. 

Alternatively, the summation procedure could be the following. Place $\mathbf{V}^{r}$ on the face $\mathcal{VBC}$ and place $\mathbf{V}^{s}$ on the contiguous faces: first on $\mathcal{VAB}$ (to count the edge $\mathcal{VB}$) and then on the face $\mathcal{VCD}$ (to account for $\mathcal{VC}$). That is, fix index $r$ and sum over $s$. We still need to consider the contribution of the edges $\mathcal{VD}$ and $\mathcal{VA}$. We then take $\mathbf{V}^{r}$ on $\mathcal{VDA}$ and place $\mathbf{V}^{s}$ on $\mathcal{VCD}$ (to count $\mathcal{VD}$); after that $\mathbf{V}^{s}$ is placed on $\mathcal{VAB}$  ( to account for $\mathcal{VA}$). When we changed the position of $\mathbf{V}^{r}$ we performed the summation over $r$. This completes te sum of both indexes.

This method described in the two paragraphs above are evidently equivalent: in both cases we add the contribution of all the edges in Fig.~\ref{FigPiramide}. This explains the prescription in Eq. (\ref{quebra soma}).

After those remarks, we conclude that Eq.~(\ref{soma deltaE Lp}) can be cast into the form :%
\begin{equation}
\sum\limits_{p}\delta\varepsilon_{p}L_{p}=\sum\limits_{p}\left[  \sum_{s}%
V_{s}^{\mu}\delta V_{s}^{\nu}\sum_{r}V_{\mu\nu}^{rs}L_{rs}^{p}+\sum_{r}%
V_{r}^{\nu}\delta V_{r}^{\mu}\sum_{s}V_{\mu\nu}^{rs}L_{rs}^{p}\right] \,
.\label{Sum var epsilon Lp}
\end{equation}
For orthogonality reasons, it is true that\footnote{$\mathbf{V}^{rs}$ is the tensor defined over the area formed by $\mathbf{V}^{r}$ and $\mathbf{V}^{s}$ (Fig.~{\ref{FigOrtogonais}}); $L_{rs}$ is the measure of the joint area $\left(rs\right)  $ and can be understood as $\left(  \mathbf{V}^{r}\times
\mathbf{V}^{s}\right)  $. As we have discussed, $\mathbf{V}^{rs}$ is
\textquotedblleft orthogonal\textquotedblright\ to $\left(  \mathbf{V}%
^{r}\times\mathbf{V}^{s}\right)  $; therefore, the internal product of these
quantities vanishes.}
\begin{equation}
\sum_{s}V_{\mu\nu}^{rs}L_{rs}^{p}=0 \, , \label{ortogonalidade}%
\end{equation}
For this reason, the right-hand side of Eq.~(\ref{Sum var epsilon Lp}) vanishes entirely:%
\begin{equation}
\sum\limits_{p}\delta\varepsilon_{p}L_{p}=0 \, , \label{var epsilon Lp}%
\end{equation}
as anticipated.


\subsection{Evaluating $\delta L_{p}$
\label{sec-area}}

In four dimensions, the sheaf will not be a straight (oriented) line of length
$l_{p}$. On the contrary, it will be a bidimensional surface with area $L_{p}$ (and border
$l_{p}$, for example). We are interested in obtaining the area $L_{p}$ as a function of
$l_{p}$ in order to establish the discrete version of the field equations for gravity in 4D.

\begin{figure}[h]
\begin{center}
\includegraphics[scale=0.65]{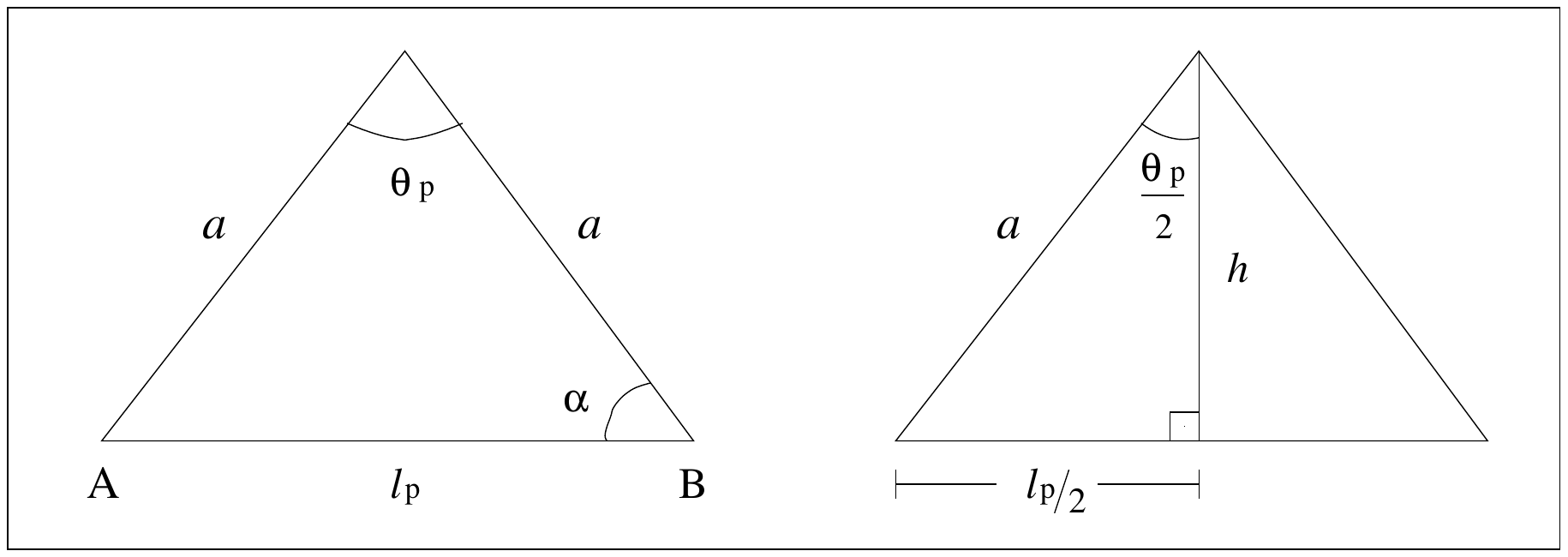}
\end{center}
\caption{Isosceles triangle of sides $a$, $a$ and $l_{p}$. $\theta_{p}$ is the
opposite angle to $l_{p}$, and $\alpha$\ the opposite angle to $a$.}%
\label{FigIsosceles}%
\end{figure}

For simplicity, let us consider the bidimensional area $L_{p}$ as the isosceles triangle
of base $l_{p}$; cf. Fig.~\ref{FigIsosceles}.


From elementary geometry, the area $L_{p}$ is 
\begin{equation}
L_{p}=\frac{1}{2}\left(  \text{base}\right)  \left(  \text{hight}\right)
=\frac{1}{2}l_{p}h~; \label{formula da area}%
\end{equation}
and the hight $h$ follows from
\begin{equation}
a^{2}=\left(  \frac{l_{p}}{2}\right)  ^{2}+h^{2} \, . \label{Pitagoras}%
\end{equation}
The side $a$ is:%
\begin{equation}
\sin\frac{\theta_{p}}{2}=\frac{\left(  l_{p}/2\right)  }{a} \Rightarrow a=\frac{l_{p}}{2}\frac{1}{\sin\frac{\theta_{p}}{2}}\, . \label{a}%
\end{equation}

Substituting (\ref{a}) into (\ref{Pitagoras}):\[
h^{2}=\left(  \frac{l_{p}}{2}\right)  ^{2}\left(  \frac{1}{\sin^{2}%
\frac{\theta_{p}}{2}}-1\right)  \, ,
\]
i.e,%
\begin{equation}
h=\frac{l_{p}}{2}\frac{\cos\frac{\theta_{p}}{2}}{\sin\frac{\theta_{p}}{2}}~.
\label{h}%
\end{equation}
Hence, the area equation (\ref{formula da area})  turns to
\begin{equation}
L_{p}=\left(  \frac{l_{p}}{2}\right)  ^{2}\cot\frac{\theta_{p}}{2} \, .
\label{Lp}%
\end{equation}


\begin{figure}[h]
\begin{center}
\includegraphics[scale=0.6]{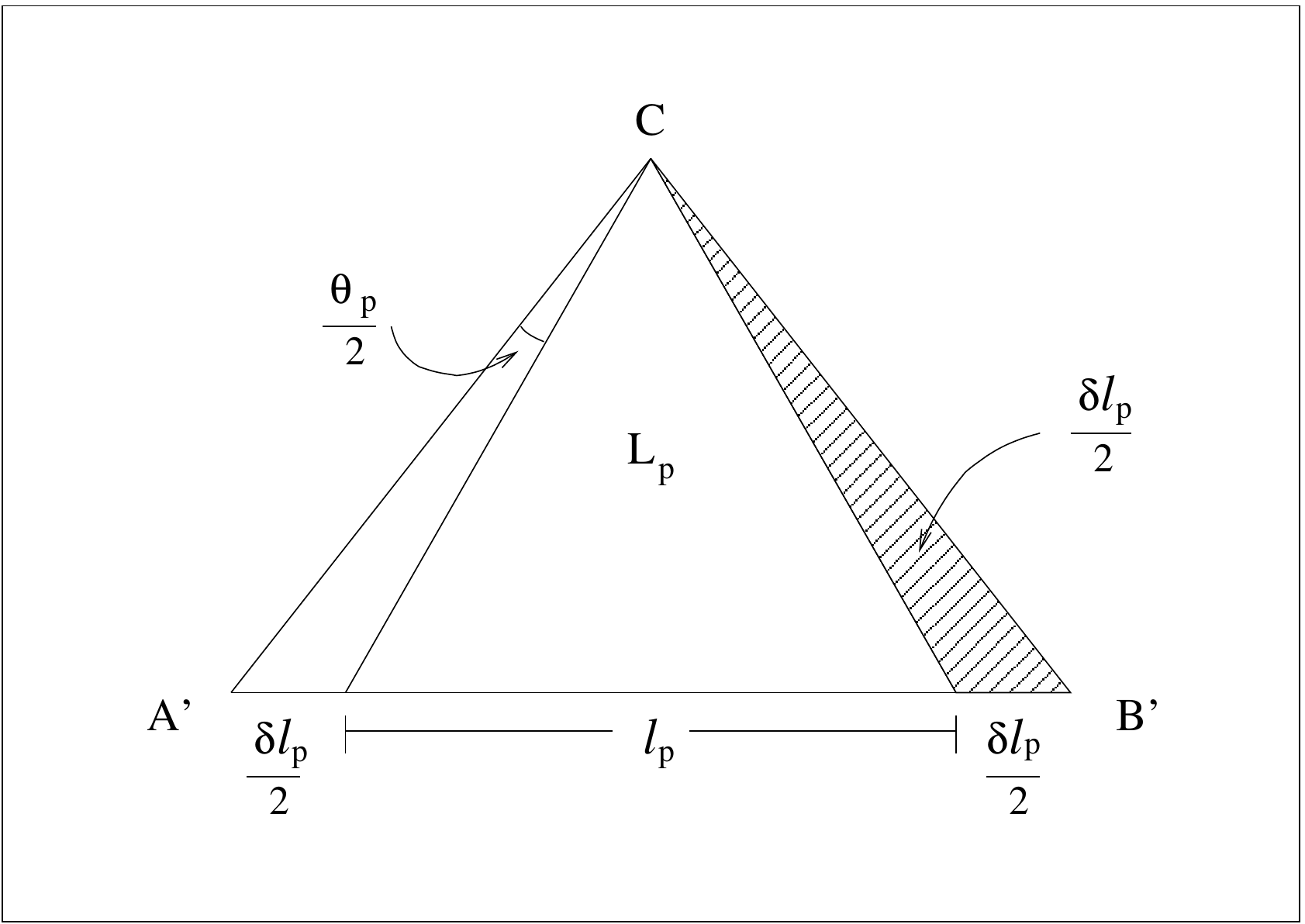}
\end{center}
\caption{Variation of the area of the isosceles triangle by increment in
$l_{p}$.}%
\label{FigArea}%
\end{figure}


If we vary $l_{p}$ in a way to increment by equal quantities $\delta l_{p}/2$
both extremities of the base, the triangle remains isosceles --- see Fig.~\ref{FigArea}. The opposite
angle to the incremented base $l_{p}^{\prime}$,
\[
l_{p}^{\prime}=\frac{\delta l_{p}}{2}+l_{p}+\frac{\delta l_{p}}{2}%
\]
increases to
\[
\theta_{p}^{\prime}=\frac{\delta\theta_{p}}{2}+\theta_{p}+\frac{\delta
\theta_{p}}{2} \, .
\]
Since the post-variation triangle remains isosceles, its area
$L_{p}^{\prime}=L_{p}+\delta L_{p}$\ is calculated by Eq.~(\ref{Lp}):
\[
L_{p}^{\prime}=\left(  \frac{l_{p}^{\prime}}{2}\right)  ^{2}\cot\left(
\frac{\theta_{p}^{\prime}}{2}\right)  =\left(  \frac{l_{p}+\delta l_{p}}%
{2}\right)  ^{2}\cot\left(  \frac{\theta_{p}+\delta\theta_{p}}{2}\right)  \, ;
\]
where
\[
\left(  l_{p}+\delta l_{p}\right)  ^{2}=l_{p}^{~2}+2l_{p}~\delta l_{p}+\delta
l_{p}^{~2}\simeq l_{p}^{~2}+2l_{p}~\delta l_{p} \, ,
\]
since second order increments are neglected. Moreover, expanding $\cot\theta
_{p}^{\prime}$ in Taylor series about $\theta_{p}/2$ yields:
\[
\cot\left(  \frac{\theta_{p}+\delta\theta_{p}}{2}\right)  =\cot\frac
{\theta_{p}}{2}+\left.  \frac{d}{d\theta_{p}}\cot\left(  \frac{\theta
_{p}+\delta\theta_{p}}{2}\right)  \right\vert _{\theta_{p}/2}\frac
{\delta\theta_{p}}{2}+\mathcal{O}\left(  \frac{\delta\theta_{p}^{~2}}{4}\right)  \, ,
\]
where
\[
\frac{d}{dx}\cot x=-\frac{1}{\sin^{2}x}=-\left(  1-\cot^{2}x\right)  \, .
\]
It then follows:
\[
L_{p}^{\prime}=\frac{\left(  l_{p}^{~2}+2l_{p}\delta l_{p}\right)  }{4}\left[
\cot\frac{\theta_{p}}{2}-\frac{1}{\sin^{2}\frac{\theta_{p}}{2}}\frac
{\delta\theta_{p}}{2}\right]  +\mathcal{O}\left(  \delta l_{p}^{~2}%
,\delta\theta_{p}^{~2}\right)  \, ,
\]
i.e.,%
\[
L_{p}+\delta L_{p}=\frac{l_{p}^{~2}}{4}\cot\frac{\theta_{p}}{2}+\frac{1}%
{2}l_{p}\delta l_{p}\cot\frac{\theta_{p}}{2}-\frac{1}{4}l_{p}^{~2}\frac
{1}{\sin^{2}\frac{\theta_{p}}{2}}\frac{\delta\theta_{p}}{2}+\mathcal{O}\left(
\delta l_{p}^{~2},\delta\theta_{p}^{~2}\right)  \, .
\]
The first term in the right-hand side of the equation above is precisely
$L_{p}$, Eq. (\ref{Lp}). It cancels out the first term in the left-hand side. 
Thus,
\begin{equation}
\delta L_{p}\simeq\frac{1}{2}l_{p}\delta l_{p}\cot\frac{\theta_{p}}{2}%
-\frac{1}{4}l_{p}^{~2}\frac{1}{\sin^{2}\frac{\theta_{p}}{2}}\frac{\delta
\theta_{p}}{2} \, .\label{var Lp var theta}%
\end{equation}

\begin{figure}[h]
\begin{center}
\includegraphics[scale=0.6]{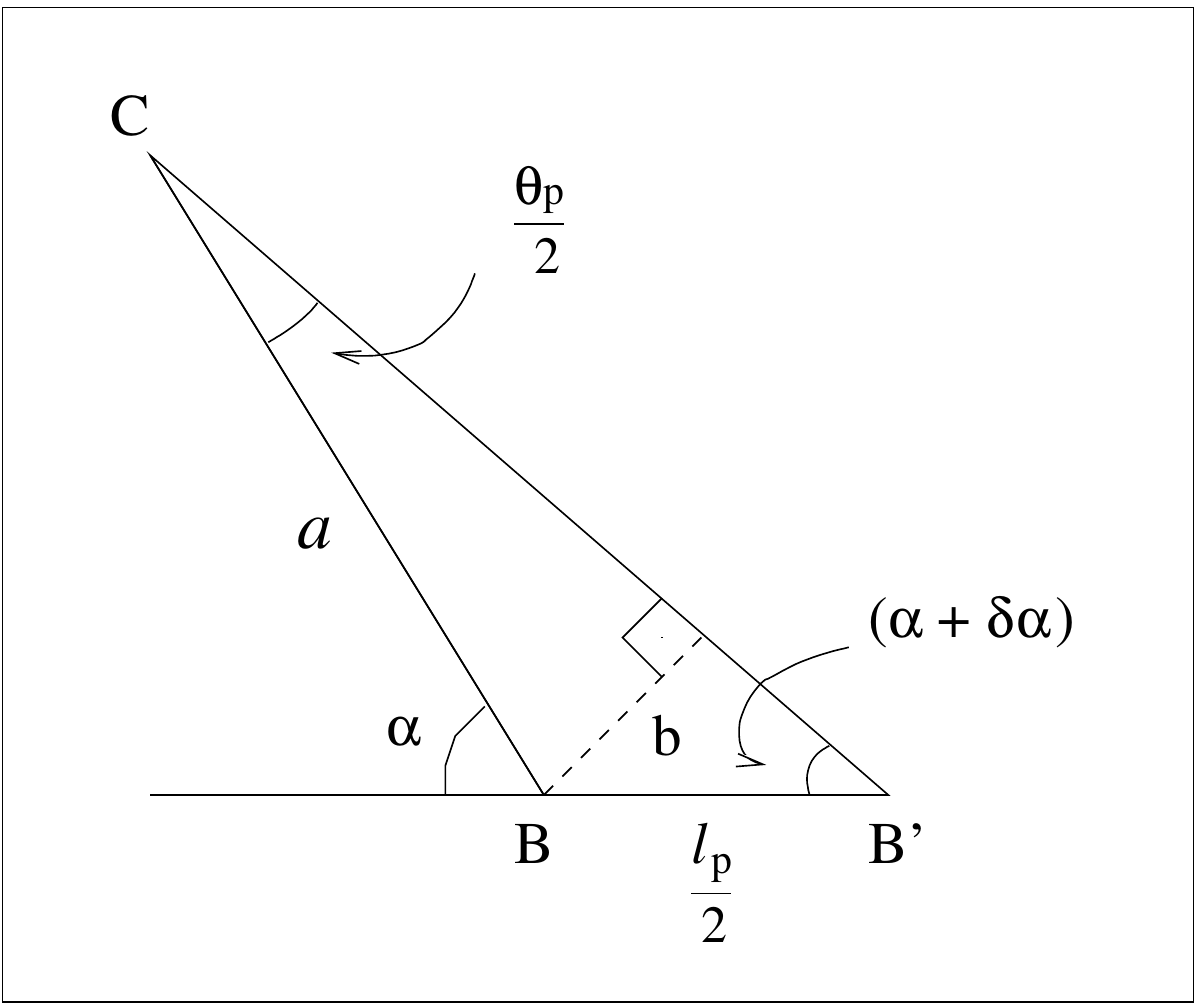}
\end{center}
\caption{Triangle $BCB'$ resulting from the increment $\delta l_{p}$ on the base of triangle $ABC$ of Fig.~\ref{FigIsosceles}. The same triangle is highlighted in Fig.~\ref{FigArea}.}%
\label{FigIncremento}%
\end{figure}

The term $\delta\theta_{p}$ in (\ref{var Lp var theta}) is of the same order 
as $\delta l_{p}$. Fig.~\ref{FigIncremento} helps us to check this out.


Based on Fig.~\ref{FigIncremento}, we write:
\begin{equation}
\sin\left(  \alpha+\delta\alpha\right)  =\frac{b}{\left(  \delta
l_{p}/2\right)  } \, . \label{sin alpha}
\end{equation}
On the other hand, from the geometry of the original isosceles triangle in Fig.~\ref{FigIsosceles}:
\[
\alpha=\frac{\pi}{2}-\frac{\theta_{p}}{2} \, .
\]
Therefore, $\delta \alpha$ reads:
\[
\delta\alpha=-\frac{\delta\theta_{p}}{2} \, .
\]
Consequently,
\begin{equation}
\sin\left(  \alpha+\delta\alpha\right)  =\sin\left[  \frac{\pi}{2}%
-\frac{\left(  \theta_{p}+\delta\theta_{p}\right)  }{2}\right]  =\cos\left(
\frac{\theta_{p}+\delta\theta_{p}}{2}\right)  \, \label{cos theta}.
\end{equation}
Equating (\ref{sin alpha}) and (\ref{cos theta}):
\[
\cos\left(  \frac{\theta_{p}+\delta\theta_{p}}{2}\right)  =\frac{b}{\left(
\delta l_{p}/2\right)  } \, ;
\]
i.e.%
\begin{equation}
b=\frac{\delta l_{p}}{2}\cos\left(  \frac{\theta_{p}+\delta\theta_{p}}%
{2}\right)  \, . \label{b}%
\end{equation}

Yet another result from Fig.~\ref{FigIncremento}:
\[
\sin\left(  \frac{\delta\theta_{p}}{2}\right)  =\frac{b}{a} \, .
\]
Substituting the expressions (\ref{a}) and (\ref{b}) for $a$ and $b$:
\[
\sin\left(  \frac{\delta\theta_{p}}{2}\right)  =\frac{\delta l_{p}}{l_{p}}%
\sin\frac{\theta_{p}}{2}\cos\left(  \frac{\theta_{p}+\delta\theta_{p}}%
{2}\right)  \, .
\]
Now, we use the fact
\[
\frac{\delta\theta_{p}}{2}\ll1
\]
to approximate
\[
\sin\left(  \frac{\delta\theta_{p}}{2}\right)  =\frac{\delta\theta_{p}}%
{2}+\mathcal{O}\left(  \frac{\delta\theta_{p}^{~3}}{8}\right)
\]
and
\[
\cos\left(  \frac{\theta_{p}+\delta\theta_{p}}{2}\right)  =\cos\left(
\frac{\theta_{p}}{2}\right)  +\mathcal{O}\left(  -\sin\frac{\theta_{p}}%
{2}~\frac{\delta\theta_{p}}{2}\right)  \, ,
\]
so that
\begin{equation}
\frac{\delta\theta_{p}}{2}\simeq\frac{\delta l_{p}}{l_{p}}\sin\frac{\theta
_{p}}{2}\cos\frac{\theta_{p}}{2} \, . \label{var theta}%
\end{equation}

Substituting (\ref{var theta}) into Eq.~(\ref{var Lp var theta}) for $\delta L_{p}$, we get:
\[
\delta L_{p}\simeq\frac{1}{2}l_{p}\delta l_{p}\cot\frac{\theta_{p}}{2}%
-\frac{1}{4}l_{p}\delta l_{p}\frac{\cos\frac{\theta_{p}}{2}}{\sin\frac
{\theta_{p}}{2}} \, .
\]
Therefore, up to order-$\left(  \delta l_{p}\right)  ^{2}$ terms, it results:%
\begin{equation}
\delta L_{p}=\frac{1}{4}l_{p}\delta l_{p}\cot\frac{\theta_{p}}{2} \, .
\label{var Lp}%
\end{equation}
This is the last ingredient we needed for writing the final version of Regge's action integral.


\subsection{The discretized version of Einstein equations
\label{sec-Einstein}}

Substituting (\ref{var epsilon Lp}) and (\ref{var Lp}) into (\ref{var I}), it follows:
\[
\delta I=\frac{1}{8\pi}\sum\limits_{p}\varepsilon_{p}\left(  \frac{1}{4}%
l_{p}\delta l_{p}\cot\frac{\theta_{p}}{2}\right)  \, .
\]

We now adopt the approximation in which all $p$ joints in the
manifold have the same length $l_{p}$ on average,
\[
l_{p}\simeq l \, .
\]
This is consistent with the simplicity hypothesis that is convenient to triangulate the manifold using polyhedrons as close to regular polyhedrons as possible. It is also consistent with the $L_p$ calculation in the previous subsection.

Accordingly, the variations with respect to $l_{p}$ should be, on average,
the same for all $p$ joints. This means:
\[
\delta l_{p}\simeq\delta l \, ,
\]
leading to
\begin{equation}
\frac{\delta I}{\delta l}=\frac{1}{32\pi}l\sum\limits_{p}\varepsilon_{p}%
\cot\frac{\theta_{p}}{2}~, \label{dIdl}
\end{equation}
since the product $l_{p}~\delta l_{p}\simeq l~\delta l$ can be taken off the sum in $p$.

The principle of minimal action manifested in Eq.~(\ref{delta I simpletica}) enforces the vanishing of Eq.~(\ref{dIdl}) which leads to Einstein equations for the discretized space:
\begin{equation}
\sum\limits_{p}\varepsilon_{p}\cot\frac{\theta_{p}}{2}=0 \, .
\label{Einstein simpletica}
\end{equation}
This result is analogous to (\ref{Einstein RG}).


\section{Final Remarks}

In this paper, we have meticulously studied the seminal work by Regge on the discretized version of general relativity. We have made an effort to bring the abstract and synthetic style of the original paper \cite{Re61} to a more down to Earth and step-by-step approach to the subject. Regge calculus is a fundamental theoretical  background to a numerical treatment of curvature and there lies the key importance of the subject. 

Regge's discretized version of gravitation has a parallel with the Quantum
Chromodynamics Theory in its attempt to perform network calculations
\cite{Smit}. Indeed, some argue that Regge calculus would be the
appropriate way to quantize gravitation, despite its limitations
\cite{Immirzi}.

Regge calculus was applied to the Schwarzschild solution and to the study of
Reissner-Nordstr\"om geometry \cite{Wh64}. The Friedmann models were
also treated in the light of this method \cite{Wh64}. Ref. \cite{Ha86} builds 
the simplex version of the action integral for higher order gravity. The construction of 
the teleparallel equivalent of Regge equations in \cite{Pe02} is another application to 
gravitation. Refs.~\cite{Mc10,Mi14} contain a formal approach to Regge calculus and further 
applications (see also references therein).

Recently, advances in computational capacity have led to a renewed interest in
Regge calculus, especially after the successes obtained within Quantum
Chromodynamics. This culminated in a new version of Regge's theory called
\emph{Dynamic Causal Triangulation} \cite{CDT}.  

Perhaps the most recent application of the subject of this paper is the
relationship between Regge calculus and the spin networks used in Loop Quantum
Gravity. In fact, it is possible to show that spin networks are a dual
representation of Regge's spacetime triangulation. For instance, in order
to picture a tetrahedron as a spin network, we use a vertex to denote the
volume and four links to represent the four faces. The value of the volume is 
given by a number in the vertex and the faces' areas are related to
four numbers (one for each link)\footnote{We refer the interested reader to
Fig.~4.3 of Ref.~\cite{IntroLQG}. See also Fig.~1.4 in the same reference.}.
In the quantized version, each area is described by an area operator, which is
closed under the $su(2)$ algebra plus a closure relation \cite{IntroLQG}. Each
vertex is expressed by a volume operator, and the spin network is the
basis which simultaneously diagonalizes both operators. Regge
calculus is very important for Loop Quantum Gravity to obtain the necessary
classical limit.

It is our hope this paper will facilitate the interested reader to enter the field.

\subparagraph{Acknowledments}

The authors are grateful to Ruben Aldrovandi, Jos\'e G. Pereira, Bruto M.
Pimentel and Te\'ofilo Vargas from IFT-Unesp (Brazil) for references and 
insightful discussions.


\end{document}